%% file: phaseTIR_arxiv_v2.tex
\documentclass[a4paper,11pt]{article}
\usepackage{jheppub}
\usepackage{dsfont}
\usepackage{amsmath,amssymb,amscd,amsfonts,mathtools}

\usepackage[toc,page]{appendix}
\usepackage{epsfig}
\usepackage{epstopdf}
\usepackage{latexsym}
\usepackage{graphicx}
\usepackage{placeins}
\usepackage{floatrow}
\usepackage{subfig}
\usepackage{caption}
\usepackage{booktabs}
\usepackage{bbm}
\usepackage{color}
\usepackage{physics}
\usepackage{stackrel}
\usepackage{tensor}
\usepackage{tikz}
\usetikzlibrary{matrix}
\usetikzlibrary{decorations.markings,calc,shapes,decorations.pathmorphing,patterns,decorations.pathreplacing}
\usetikzlibrary{positioning}
\usepackage{hyperref}

\pdfoutput=1
\makeatletter
\def\@fpheader{\relax}
\makeatother

\def\L{{\cal L}}

\def\tilde#1{{\widetilde{#1}}}

\DeclareMathSizes{10}{10}{7}{7}

\subheader{\begin{flushright}
UTWI-17-2023\\
\end{flushright}}

\title{Imprints of phase transitions on Kasner singularities}

\author{Elena Caceres,}
\author{Sanjit Shashi,}
\author{Hao-Yu Sun}
\affiliation{Theory Group, Weinberg Institute, Department of Physics, University of Texas,\\
\phantom{}\hspace{0.5cm} 2515 Speedway, Austin, Texas 78712, USA.}
\emailAdd{elenac@utexas.edu}
\emailAdd{sshashi@utexas.edu}
\emailAdd{hkdavidsun@utexas.edu}

\abstract{Under the AdS/CFT correspondence, asymptotically AdS geometries with backreaction can be viewed as CFT states subject to a renormalization group (RG) flow from an ultraviolet (UV) description towards an infrared (IR) sector. For black holes however, the IR point is the horizon, so one way to interpret the interior is as an analytic continuation to a ``trans-IR" imaginary-energy regime. In this paper, we demonstrate that this analytic continuation preserves some imprints of the UV physics, particularly near its ``endpoint" at the classical singularity. We focus on holographic phase transitions of geometric objects in round black holes. We first assert the consistency of interpreting such black holes, including their interiors, as RG flows by constructing a monotonic $a$-function. We then explore how UV phase transitions of entanglement entropy and scalar two-point functions, each of which are encoded by bulk geometry under the holographic mapping, are connected to the structure of the near-singularity geometry, which is characterized by Kasner exponents. Using 2d holographic flows triggered by relevant scalar deformations as test beds, we find that the 3d bulk's near-singularity Kasner exponents can be viewed as functions of the UV physics precisely when the deformation is nonzero.}
\begin{document}
\maketitle
\vfill
\pagebreak

\section{Introduction} 

A renormalization group (RG) flow is a contour in the space of QFT couplings that describes the coarse graining of a high-energy (UV-complete) theory to a low-energy (IR) description \cite{Wilson:1973jj,Polchinski:1983gv}. We integrate out all high-energy modes above some cutoff energy to get an effective QFT. The flow itself consists of all such effective theories and is thus parameterized by the cutoff. Furthermore, both the IR and UV theories are conformal-invariant fixed points of the flow and so are conformal field theories (CFTs). Most importantly, the IR theory may have imprints of the UV physics as remnants of the integration procedure, even though the IR is mostly insensitive to the UV.

A guiding principle of modern quantum gravity research is the AdS/CFT correspondence \cite{Maldacena:1997re} equating weakly coupled ``bulk" gravity on $(d+1)$-dimensional anti-de Sitter (AdS) space to strongly coupled CFT physics on the $d$-dimensional boundary of AdS. AdS/CFT describes a ``holographic" class of RG flows \cite{Balasubramanian:1999jd}. The energy cutoff is associated with the radial extra dimension of the bulk theory, with the UV fixed point being the boundary CFT, so this RG flow essentially describes the emergence of holographic bulk spacetime. The flow dynamics are encoded by the gravitational theory \cite{deBoer:1999tgo,deBoer:2000cz,Bianchi:2001kw,Fukuma:2002sb,Papadimitriou:2004ap,Papadimitriou:2005ii}, which must include a bulk matter sector dual to some relevant deformation of the UV theory.\footnote{Other types of deformations (irrelevant, marginal) are possible to implement holographically, such as the $T\bar{T}$ deformation \cite{Zamolodchikov:2004ce,Smirnov:2016lqw,Cavaglia:2016oda}. However, these require introducing a finite bulk cutoff by hand \cite{McGough:2016lol,Guica:2019nzm,Iliesiu:2020zld,Ebert:2022gyn,Ebert:2022ehb}. We focus on relevant deformations so as to have geometric flows with both asymptotic UV and IR regions.} As the bulk theory is weakly coupled, it is a tractable setting in which one can study imprints of UV physics in the IR sector.

Gravitational theories typically feature black holes, and these are dual to states of a canonical ensemble at fixed temperature. For example, large AdS black holes are dual to thermal CFT states \cite{Witten:1998zw,Hubeny:2014bla,hartmanhepQG}. From the perspective of holographic RG flow, such a black hole with matter-induced backreaction describes an RG flow from a UV thermal state towards an IR fixed point associated with the horizon. However, the classical geometry does not stop at the horizon; there is also an interior region in which the radial dimension is timelike, rather than spacelike. Nonetheless, we may \textit{insist} that the black-hole interior is also part of an RG flow. We would then claim that the interior corresponds to a ``trans-IR" part of the flow \cite{Frenkel:2020ysx,Caceres:2022smh} defined as an analytic continuation of the exterior RG flow to imaginary energies. It is further natural to infer that the UV physics leaves an imprint on the trans-IR regime. Exploring such imprints is the goal of this paper.

We emphasize that the trans-IR picture is a working interpretation rather than a firm, established statement. In particular, it is not clear what the actual ``points" along this interior RG flow (reached by integrating out all real-energy modes and integrating \textit{in} ``imaginary-energy" modes) would represent in field-theoretic terms, although some progress has been made by \cite{Hartnoll:2022snh}. While we do not directly tackle this issue in this paper, the study herein is meant to provide some internal consistency to the trans-IR perspective of the black-hole interior as a coarse-graining flow. To this end, we first build on previous work \cite{Caceres:2022smh,Caceres:2022hei} exploring the existence of monotonic ``degree-of-freedom-counting" functions.

Furthermore, we explore how phase transitions in the UV are encoded by the trans-IR part of the flow. We focus on the following two holographic phase transitions in black holes with spherical horizons:
\begin{itemize}
\item[(1)] the deviation of a boundary subregion's entanglement from the Araki--Leib bound \cite{Araki:1970ba} described by the holographic entanglement plateaux of \cite{Hubeny:2013gta}; and

\item[(2)] the connected/disconnected phase transition of a thermal two-point function of a heavy scalar operator $\mathcal{O}_{\text{H}}$ corresponding to a transition of its holographic geodesic approximation \cite{Banks:1998dd,Balasubramanian:1999zv}.
\end{itemize}
The question of how UV phase transitions imprint upon the interior geometry has been tackled in the planar holographic superconductor \cite{Hartnoll:2008kx} by \cite{Hartnoll:2020fhc}, which found that approaching the condensate temperature from below coincides with rapid, fractal-like fluctuations in the scaling of space deep inside the black-hole geometry. Indeed, one may study the imprints of phase transitions on the trans-IR regime across a wide variety of AdS/CMT models with specific matter content by examining the evolution of black-hole interiors \cite{Mansoori:2021wxf,Liu:2021hap,Cai:2021obq,Liu:2022rsy,An:2022lvo,Mirjalali:2022wrg,Sword:2022oyg,Gao:2023zbd}. However, we are interested in more universal imprints arising from phase transitions that are not contingent on having a particular type of matter field. Any bulk theory that includes round black holes as classical solutions will also accommodate the above phase transitions.

Black-hole solutions to Einstein gravity with no matter have interior geometries described by finely tuned Kasner cosmologies \cite{Kasner:1921zz}. However, we find that turning on a scalar deformation in the UV (and thus a matter field in the bulk) changes the story and allows UV physics to imprint upon the near-singularity geometry. Specifically, the deep-interior Kasner cosmologies develops a nontrivial relationship with parameters that characterize the above geometric transitions.

\subsection{Horizon topology and phase transitions}

Throughout this paper, we restrict ourselves to backreacting AdS black holes of the form
\begin{equation}
ds^2 = \frac{\ell^2}{r^2}\left[-e^{-\chi(r)} F(r) dt^2 + \frac{dr^2}{F(r)} + \ell^2 d\Sigma_{d-1,k}^2\right],\label{metTop}
\end{equation}
where $t \in \mathbb{R}$ and $r > 0$, with the conformal boundary being at $r = 0$. $\chi$ is a real function, and $F$ has a simple root at $r = r_{\text{h}}$ (defining the horizon). $d\Sigma_{d-1,k}$ is a unit line element describing the topology of the horizon. As in the pure gravity \cite{Birmingham:1998nr}, this line element is labeled by $k = \{-1,0,1\}$, respectively describing hyperbolic, planar, and spherical horizons:
\begin{equation}
d\Sigma_{d-1,k}^2 = \begin{cases}
d\mathbb{H}_{d-1}^2,&\text{if}\ k = -1,\\
d\hat{x}_{d-1}^2,&\text{if}\ k = 0,\\
d\Omega_{d-1}^2,&\text{if}\ k = 1.
\end{cases}\label{topk}
\end{equation}
The study of the planar case was initiated by \cite{Frenkel:2020ysx}, with an associated monotonic $a$-function having been identified in \cite{Caceres:2022smh}. In this paper, it is the spherical case that is of interest.

The main underlying difference between the spherical case and the others is that spheres are compact. Owing to this, round black holes accommodate various phase transitions of bulk geometric objects, which in turn are interpreted as phase transitions of boundary CFT quantities. In the other topological cases, the phase transitions in which we are primarily interested either become much simpler or go away.

\subsection{A brief comment on the UV state}

If we organize the CFT states into a canonical ensemble, there are generically (in $d > 2$) three types of bulk states at each temperature: a large black hole, a small black hole, and a thermal gas of gravitons \cite{Witten:1998zw,Hubeny:2014bla,hartmanhepQG}. The small black hole typically never dominates the canonical partition function and is even thermodynamically unstable,\footnote{However, this does not mean small black holes have no macroscopic thermodynamic effects. For example, they are actually responsible for the peak of the bulk viscosity-to-entropy ratio near the critical temperature in confining large-$N_c$ gauge theories \cite{Gursoy:2008za,Gursoy:2010fj}.} but the other two generally exchange dominance at a Hawking--Page transition temperature \cite{Hawking:1982dh}. Therefore, while all of these states are thermal states, the large black hole is the dominant thermal state only above the Hawking--Page temperature.

Our focus in this paper is on the near-singularity structure of black holes. From the holographic-RG perspective, we should thus be careful to say that the interior is only the trans-IR flow from the dominant UV thermal state when we are looking at a large black hole on one side of an assumed Hawking--Page transition. However, we can always say that the interior represents a trans-IR flow from \textit{some} thermal state.

In this work, we do not study the Hawking--Page transition in the presence of matter. However, it should be possible to pin down where this transition happens (cf. \cite{Gursoy:2010jh}). We leave this to future work.

\subsection{Outline}

In Section \ref{sec:roundBH}, we argue for the irreversibility of the flow from the UV of a generic round black hole to its IR, as well as the irreversibility of this flow's trans-IR analytic continuation towards the classical singularity. As in earlier AdS/CFT literature \cite{Alvarez:1998wr,Freedman:1999gp,Myers:2010xs,Myers:2010tj}, we assume the null energy condition to construct a monotone that counts the degrees of freedom along the flow. Our statements in this section are analogous to those of the planar case \cite{Caceres:2022smh}.

In Section \ref{sec:data}, we elaborate on what specifically we mean by ``data" in the context of holographic RG flows triggered by a minimal class of scalar deformations. While these flows do not feature the types of phase transitions often seen in AdS/CMT, they still support round black holes and allow geometric phase transitions, so they are sufficient for our purposes here. We will also show how the near-singularity geometry is a function of the temperature only when a deformation is turned on.

After establishing the basic machinery---the equations of motion, the list of UV data, and the construction of ``round" Kasner universes \cite{Kasner:1921zz} describing the near-singularity geometry---we then discuss the entanglement plateaux (Section \ref{sec:plateaux}) and the phase structure of the heavy thermal (with respect to the black-hole state) two-point function (Section \ref{sec:thermal2pt}) in these scalar flows and how they imprint upon the near-singularity geometry.

\section{Holographic $a$-theorem for round black holes}\label{sec:roundBH} 

In the framework of RG flows, it is natural to ask how to count the degrees of freedom to quantify the idea of coarse graining a theory. In principle, this should be described by a function which decreases monotonically as we flow from the UV to the IR, since flowing in this manner corresponds to integrating out degrees of freedom \cite{Wilson:1973jj,Polchinski:1983gv}. A flow with such a function is ``irreversible."

In general QFT, the seminal work on this front is Zamolodchikov's $c$-theorem \cite{Zamolodchikov:1986gt} stating that 2d RG flows of unitary, Lorentz-invariant theories feature a monotonic $c$-function that coincides with the central charges at the RG fixed points. Cardy's conjectured extension to 4d \cite{Cardy:1988cwa} (and any even number of dimensions, for that matter), the $a$-theorem, asserts the existence of an $a$-function that coincides with the $A$-type trace anomaly coefficients at the fixed points. The $a$-theorem in 4d has been proven for nonholographic flows \cite{Komargodski:2011vj}. However, holographic flows are nice in part because the $a$-theorem can be extended to and proven in any number of dimensions \cite{Freedman:1999gp,Myers:2010xs,Myers:2010tj} assuming reasonable energy constraints on the bulk matter, although the field-theoretic interpretation of the odd-dimensional holographic $a$-function is not connected to an anomaly but rather entanglement entropy.

If the black-hole interior is to be interpreted as an analytically continued RG flow, then the ability to count degrees of freedom should extend to the trans-IR regime. This has been done for flat black holes in Einstein gravity \cite{Caceres:2022smh,Caceres:2022hei}. Here, we construct the $a$-function of a round black hole in Einstein gravity. Our goal in this endeavor is to perform a preliminary consistency check of interpretation of the interior geometry as a coarse-graining flow, even though ``energy" is imaginary.

As a caveat, we note that such a monotone is a rather coarse consistency check and its mere existence does not address questions surrounding what it physically means to flow along imaginary energies on the level of field theory. We leave addressing this point to future work, for now simply assuming that the existence of a monotone serves as reasonable evidence for \textit{some} RG-flow interpretation of the interior.

\subsection{$a$-functions from the null energy condition}

The trace anomaly was first realized holographically by \cite{Henningson:1998gx} in Einstein gravity,
\begin{equation}
G_{\mu\nu} - \frac{d(d-1)}{2\ell^2} g_{\mu\nu} = 0,
\end{equation}
where $\ell$ is the curvature radius. The trace anomaly coefficient $a_*$ goes as $\ell^{d-1}$, and in keeping with the normalization conventions of \cite{Myers:2010tj,Myers:2010xs} it is
\begin{equation}
a_* = \frac{\pi^{d/2}}{\Gamma\left(\frac{d}{2}\right)\ell_{\text{P}}^{d-1}}\ell^{d-1}.\label{holoTrace}
\end{equation}
For example, for $d = 2$ (where $\ell_{\text{P}} = 8\pi G_{\text{N}}$) we reproduce the usual 2d trace anomaly coefficient with the Brown-Henneaux central charge $c_{\text{BH}} = \frac{3\ell}{2G_{\text{N}}}$ \cite{Brown:1986nw},
\begin{equation}
a_*|_{d=2} = \frac{\ell}{8 G_{\text{N}}} = \frac{c_{\text{BH}}}{12}.
\end{equation}
Holographic $a$-functions in Einstein gravity\footnote{One can also work in higher-derivative theories---\cite{Myers:2010tj,Myers:2010xs} does Gauss--Bonnet for example---but note that the trace anomaly and the resulting $a$-function change.} can be constructed by using $a_*$ as a starting point \cite{Freedman:1999gp}. Basically, we take gravity sourced by matter such that there is a relevant deformation at the boundary triggering an RG flow. The bulk equation of motion is then
\begin{equation}
G_{\mu\nu} - \frac{d(d-1)}{2\ell^2} g_{\mu\nu} = \ell_{\text{P}}^{d-1}T_{\mu\nu},\label{einstein}
\end{equation}
where $\ell_{\text{P}}$ is the Planck length and $T_{\mu\nu}$ is the stress tensor. We then impose a ``radial" null energy condition on matter,\footnote{While one may impose the full null energy condition, taking it to hold only for radially directed null vectors is sufficient. Heuristically, this is because the radial direction is privileged in the language of RG flow as the ``direction of energy," so only radial null energy conditions may be interpreted as imposing positivity ``along the flow."}
\begin{equation}
T_{\mu\nu}k^\mu k^\nu \geq 0,\label{necGeneral}
\end{equation}
where $k^\mu$ is some null vector pointing at least partially in the bulk radial direction orthogonal to the conformal boundary. Given a domain-wall ansatz \cite{Freedman:1999gp} for the spacetime line element such as
\begin{equation}
ds^2 = g_{ij}(\rho,\vec{x})dx^i dx^j + d\rho^2,\ \ i,j = 0,...,d-1,
\end{equation}
and that we assume to be asymptotically AdS, the inequality \eqref{necGeneral} is then used to prove the monotonicity of a combination of metric functions that reduces to $a_*$ at the conformal boundary.

We can also work in the reverse order by starting with \eqref{necGeneral} on some metric and engineering a monotone which coincides with the holographic trace anomaly coefficient. This latter systematic approach is useful for constructing candidate $a$-functions for spacetimes involving multiple metric functions, as seen in \cite{Caceres:2022hei}.

Let us take this approach to construct a monotone for spherically-symmetric round black holes. The domain-wall ansatz which foliates the bulk into cylindrical slices is
\begin{equation}
ds^2 = e^{2A(\rho)}\left[-f(\rho)^2 dt^2 + \ell^2 e^{2\mathcal{R}(\rho)} d\Omega_{d-1}^2\right] + d\rho^2.\label{bhDomain}
\end{equation}
Here, $(t,\Omega_{d-1})$ parameterize the transverse $\mathbb{R} \times S^{d-1}$ slices, and $\rho > 0$ is the radial coordinate. $A$, $\mathcal{R}$, and $f$ are metric functions that for $\rho \gg \ell$ asymptote to
\begin{equation}
A(\rho) \sim \log\cosh\left(\frac{\rho}{\ell}\right),\ \ \mathcal{R}(\rho) \sim \log\tanh\left(\frac{\rho}{\ell}\right),\ \ f(\rho) \sim 1.\label{asymptBehaviorBH}
\end{equation}
This is the condition that the geometry is asymptotically AdS$_{d+1}$ near the conformal boundary, which we recall as being the UV region. Additionally, we assume that $f(\rho) > 0$ in the exterior.

By uniformly setting $f(\rho) = 1$, we get a class of metrics describing deformations of global AdS. The warm-up construction of $a$-functions in such geometries is the subject of Appendix \ref{app:pureAdS}. In the main text, we instead focus on black-hole geometries, for which we take $f$ to have a simple root at $\rho = 0$ \cite{Hartman:2013qma}. This condition corresponds to the presence of a black-hole horizon with temperature
\begin{equation}
T_{\text{h}} = \frac{e^{A(0)}f'(0)}{2\pi}.\label{hawkTemp}
\end{equation}
While $\rho > 0$ covers the exterior, the interior is charted by analytically continuing the $\rho$ and $t$ coordinates as
\begin{equation}
\rho = i\kappa,\ \ t = t_{\text{I}} - \text{sgn}(t_{\text{I}})\frac{i\gamma}{2T_{\text{h}}},\label{analyticcont}
\end{equation}
where $\kappa > 0$, $t_{\text{I}} \in \mathbb{R}$, and $\gamma \in \mathbb{Z} + \frac{1}{2}$ \cite{Hartman:2013qma}.

We are now ready to implement the scheme for constructing holographic $a$-functions from the radial null energy condition, as discussed in \cite{Caceres:2022hei}. Much of the technical details are similar to those of the case without a black hole in Appendix \ref{app:pureAdS}. As we go into more detail there, we will be sparse on the technical details in the main text in the interest of brevity. Furthermore, we emphasize that the function resulting from the scheme will be guaranteed to be a monotone in the exterior but not the interior, so the proof of monotonicity in the interior is left to Section \ref{subsec:proofInt}.

First, we consider the null vector
\begin{equation}
k^{\mu} = e^{-A(\rho)} \delta_t^\mu + f(\rho)\delta_\rho^\mu.
\end{equation}
This vector is regular everywhere. If we contract the outer product of two such null vectors against the stress tensor computed by plugging the domain-wall metric \eqref{bhDomain} into the Einstein equations \eqref{einstein}, we get
\begin{equation}
T_{\mu\nu}k^\mu k^\nu = \frac{(d-1)}{\ell_{\text{P}}^{d-1}}\left[A'(\rho) + \mathcal{R}'(\rho)\right]^2 e^{\mathcal{R}(\rho)} f(\rho) \times \left(\frac{d}{d\rho}\left[\frac{e^{-\mathcal{R}(\rho)} f(\rho)}{A'(\rho) + \mathcal{R}'(\rho)}\right]\right).
\end{equation}
The coefficients in front of the derivative factor are manifestly positive in the exterior. Thus if we define the candidate $a$-function
\begin{equation}
\mathfrak{a}(\rho) \equiv \left[\frac{e^{-\mathcal{R}(\rho)} f(\rho)}{A'(\rho) + \mathcal{R}'(\rho)}\right]^{d-1},\label{aCandBH}
\end{equation}
then we have that $\mathfrak{a}$ is monotonic (i.e., its $\rho$-derivative is positive) if
\begin{equation}
\left[\frac{e^{-\mathcal{R}(\rho)} f(\rho)}{A'(\rho) + \mathcal{R}'(\rho)}\right]^{d-2} \stackrel{?}{>} 0.
\end{equation}
It is sufficient to show that $A'(\rho) + \mathcal{R}'(\rho) > 0$ for $\rho > 0$, since both $e^{\mathcal{R}}$ and $f$ are positive on the exterior. Indeed, this is always true given the asymptotic behavior of the metric functions \eqref{asymptBehaviorBH} and assuming their analyticity. The proper argument, as utilized in \cite{Caceres:2022hei} and Appendix \ref{app:pureAdS}, involves a proof by contradiction in which we explicitly exploit analyticity to show that
\begin{equation}
\exists \rho = \rho_*\ \ \text{such that}\ \ A'(\rho_*) + \mathcal{R}'(\rho_*) = 0 \implies \left.T_{\mu\nu}k^\mu k^\nu\right|_{\rho = \rho_* + \epsilon} < 0,
\end{equation}
for some small $\epsilon > 0$. In other words, we find an approximate expression for the null energy condition in an $\epsilon$-neighborhood of a posited root $\rho = \rho_*$ that is manifestly negative.\footnote{For the black hole, this approximation has a factor of $f^2$ relative to the expression obtained from deformations of global AdS \eqref{approxGlobal}. In the exterior, this factor is positive and thus does not impact the argument.}

So, \eqref{aCandBH} is a monotone in the exterior due to the null energy condition. Noting that the physical $a$-function should coincide with the holographic trace anomaly coefficient \eqref{holoTrace}, we ultimately write it as
\begin{equation}
a_{\text{h}}(\rho;S^{d-1}) = \frac{\pi^{d/2}}{\Gamma\left(\frac{d}{2}\right) \ell_{\text{P}}^{d-1}} \left[\frac{e^{-\mathcal{R}(\rho)} f(\rho)}{A'(\rho) + \mathcal{R}'(\rho)}\right]^{d-1}.\label{aFuncDW}
\end{equation}
The $\text{h}$ subscript is to stress that this $a$-function is defined from geometries with horizons. The $S^{d-1}$ argument is to emphasize that the dual CFT states are defined on the $(d-1)$-sphere, as opposed to being on $(d-1)$-dimensional flat space like in \cite{Caceres:2022smh}.

\subsection{Monotonicity in the interior}\label{subsec:proofInt}

The domain-wall ansatz \eqref{bhDomain} is useful for constructing $a$-functions which are manifestly monotonic in black-hole exteriors. Additionally, the smooth radial extra dimension $\rho$ is cleanly identified with the energy scale. However, to cover the black-hole interior in these coordinates, we must analytically continue $\rho$ to imaginary values. This gives rise to ambiguities when checking the monotonicity of our proposed $a$-function in the interior.

A workaround is to incorporate the factor of $i$ into the metric through a coordinate transformation of \eqref{bhDomain} to a warped Schwarzschild coordinate frame,
\begin{equation}
ds^2 = \frac{\ell^2}{r^2}\left[-e^{-\chi(r)}F(r) dt^2 + \frac{dr^2}{F(r)} + \ell^2 d\Omega_{d-1}^2\right],\label{eq:metric}
\end{equation}
where $r \in \mathbb{R}$, with $r = 0$ being the conformal boundary and $r = \infty$ being the singularity. The metric functions $\chi,F$ are analytic, and $F$ has a simple root at $r = r_{\text{h}}$. We may compute the horizon temperature in terms of these metric functions to be\footnote{Note that $F'(r_{\text{h}}) < 0$, so the temperature depends on its absolute value.}
\begin{equation}
T_{\text{h}} = \frac{e^{-\chi(r_{\text{h}})/2} |F'(r_{\text{h}})|}{4\pi}.\label{hawkTemp2}
\end{equation}
The coordinate transformation $\rho = \rho(r)$ which makes \eqref{bhDomain} assume this form is
\begin{equation}
\rho(r_{\text{h}}) = 0,\ \ d\rho = -\frac{\ell\,dr}{r\sqrt{F(r)}},\ \ e^{A(\rho) + \mathcal{R}(\rho)} = \frac{\ell}{r},\ \ \frac{f(\rho)}{e^{\mathcal{R}(\rho)}} = e^{-\chi(r)/2}\sqrt{F(r)}.\label{coordinateTrans}
\end{equation}
In domain-wall coordinates, monotonicity along the full flow is the statement that
\begin{align}
\text{Exterior:}&\ \ \frac{da_{\text{h}}}{d\rho} > 0,\label{extCond}\\
\text{Interior:}&\ \ \frac{da_{\text{h}}}{d\kappa} < 0,\label{intCond}
\end{align}
where we recall that $\kappa = -i\rho$ by \eqref{analyticcont}. We already know that \eqref{aFuncDW} satisfies the exterior condition \eqref{extCond}, but we need to also show that it satisfies the interior condition \eqref{intCond}. We do so by computing the derivative along $r$ in \eqref{eq:metric} and employing the chain rule. So, the first step is to use \eqref{coordinateTrans} on \eqref{aFuncDW} to write the $a$-function as a function of $r$.
\begin{equation}
a_{\text{h}}(r;S^{d-1}) = \frac{\pi^{d/2}}{\Gamma\left(\frac{d}{2}\right)\ell_{\text{P}}^{d-1}} \left[\ell e^{-\chi(r)/2}\right]^{d-1} = a_* e^{-(d-1)\chi(r)/2}.\label{afuncrRound}
\end{equation}
Then, we may write
\begin{align}
\frac{da_{\text{h}}}{d\kappa}
&= \frac{d\rho}{d\kappa}\frac{dr}{d\rho}\frac{da_{\text{h}}}{dr}\nonumber\\
&= -\frac{r\sqrt{|F(r)|}}{\ell}\left[\frac{(d-1)\chi'(r)}{2} a_* e^{-(d-1)\chi(r)/2}\right]\nonumber\\
&= -\left[\frac{(d-1) r \sqrt{|F(r)|}}{2\ell} a_{\text{h}}(r;S^{d-1})\right]\chi'(r).
\end{align}
In the last line, the factor in square brackets is manifestly positive in the interior. Additionally, from the radial null vector
\begin{equation}
k^\mu = e^{\chi(r)/2}\delta_t^\mu - F(r) \delta_r^\mu,
\end{equation}
the null energy condition implies that
\begin{equation}
T_{\mu\nu}k^\mu k^\nu = \frac{(d-1)F(r)^2}{2r}\chi'(r) > 0 \implies \chi'(r) > 0.
\end{equation}
So, when analytically continued to the interior, the $a$-function \eqref{aFuncDW} still decreases monotonically with $\kappa$, i.e., as we flow towards the singularity. In other words, monotonicity is preserved both in the exterior and in the interior of the round black hole.

\subsection{Sanity checks of the $a$-function}

There are some simple sanity checks that we may perform on the holographic $a$-function to ensure that it is consistent with the interpretation of holographic RG flow. We do these now.

\paragraph{Constant for vacuum solutions}

As gravitational dynamics corresponds to RG flow dynamics, a lack of backreaction due to matter should correspond to a theory which does not flow away from the UV at all. This can be checked holographically by plugging the vacuum solution into our $a$-function and seeing if it uniformly evaluates to the holographic trace anomaly coefficient \eqref{holoTrace}.

For the round black hole $a$-function, this is most simply done in the $r$ coordinate. There, we know the form of the vacuum solution analytically:
\begin{equation}
\chi(r) = 0,\ \ F(r) = \frac{r^2}{\ell^2} + 1 - \left(\frac{r}{r_{\text{h}}}\right)^{d}\left(\frac{r_{\text{h}}^2}{\ell^2} + 1\right).\label{vacMet}
\end{equation}
If we plug this into \eqref{afuncrRound}, we find that
\begin{equation}
\left.a_{\text{h}}(r;S^{d-1})\right|_{\chi(r) = 0} = a_*.
\end{equation}

\paragraph{Stationary at the horizon}

The horizon should correspond to an IR fixed point. This means that the $a$-function should be stationary when evaluated at the horizon radius. This is again confirmed rather simply in the $r$ coordinate. While the $r$-derivative,
\begin{equation}
\frac{da_{\text{h}}}{dr} = -\frac{(d-1)}{2}a_{\text{h}}(r;S^{d-1})\chi'(r),
\end{equation}
is regular at the horizon (by the analyticity of the metric function $\chi$), the derivative along the flow requires multiplication by the Jacobian,
\begin{equation}
\frac{dr}{d\rho} = -\frac{r}{\ell}\sqrt{F(r)}.
\end{equation}
$F$ vanishes identically at $r = r_{\text{h}}$, so $a_{\text{h}}$ is stationary here.

\section{Scalar flows and data}\label{sec:data}

In principle, one can write ``initial data" characterizing the beginning of an RG flow from a UV fixed point. Such data consists of dimensionless combinations of scales coming from both the fixed-point theory itself and the relevant deformation which triggers the flow. Through flow's dynamics, this initial data will be related to the ``final data" characterizing the end of the flow (conventionally the IR). Our goal is to connect UV data to trans-IR data extracted from the near-singularity classical geometry, as in for example \cite{Frenkel:2020ysx,Hartnoll:2020fhc}.

For concreteness, let us restrict to RG flows triggered by a single-trace scalar deformation $\int \phi_0 \mathcal{O}$, where $\phi_0$ is the source and $\mathcal{O}$ is a relevant scalar operator in the CFT. Such flows have a place in the literature as simple toy models of holographic renormalization \cite{Kiritsis:2016kog,Gursoy:2018umf,Elander:2022ebt}. This relevant deformation may be realized holographically by a bulk matter sector consisting solely of a scalar; for further simplicity, we take this to be a Klein--Gordon scalar. The full theory we consider is then
\begin{equation}
I[g,\Phi] = \frac{1}{2\ell_{\text{P}}^{d-1}} \int \sqrt{-g} \left[R + \frac{d(d-1)}{\ell^2} - \frac{1}{2} \left(\nabla_\alpha \Phi \nabla^\alpha \Phi + m^2 \Phi^2\right)\right].\label{actscalar}
\end{equation}
The classical bulk equations of motion are thus
\begin{align}
G_{\mu\nu} - \frac{d(d-1)}{2\ell^2}g_{\mu\nu}
&= \frac{1}{2}\left[\nabla_\mu \Phi \nabla_\nu \Phi - \frac{1}{2}g_{\mu\nu} \left(\nabla_\alpha \Phi \nabla^\alpha \Phi + m^2 \Phi^2\right)\right],\label{einsteinScalar}\\
\left(\nabla_\alpha \nabla^\alpha - m^2\right) \Phi &= 0.\label{kleinGordon}
\end{align}
These equations generically determine the dynamics of scalar flows identified with AdS geometries. However, to concretely study such flows, it helps to specify particular ans\"atze for the field and the metric. For now, we will use the metric ansatz \eqref{eq:metric} (rewritten below) and a radial ansatz for $\Phi$,
\begin{equation}
ds^2 = \frac{\ell^2}{r^2} \left[-e^{-\chi(r)} F(r) dt^2 + \frac{dr^2}{F(r)} + \ell^2 d\Omega_{d-1}^2\right],\ \ \Phi = \phi(r).\label{eq:metric2}
\end{equation}
For solutions of this form, \eqref{einsteinScalar} and \eqref{kleinGordon} reduce to three independent ordinary differential equations:
\begin{align}
&\phi'' + \left(\dfrac{F'}{F} - \dfrac{d-1}{r} - \dfrac{\chi'}{2}\right)\phi' + \dfrac{\Delta(d-\Delta)}{r^2 F}\phi = 0,\label{eomround1}\\
&\chi' - \dfrac{2F'}{F} - \dfrac{\Delta(d-\Delta) \phi^2}{(d-1) r F} - \dfrac{2d}{rF} + \dfrac{2d}{r} - \dfrac{2(d-2) r}{\ell^2 F}  = 0,\label{eomround2}\\
&\chi' -  \dfrac{r}{d-1}(\phi')^2 = 0,\label{eomround3}
\end{align}
The only difference from the equations of motion for a flat topology is the last term in \eqref{eomround2}. This term reflects the topology of the horizon in our metric ansatz.

\subsection{UV data}

First, we describe the initial UV data for these scalar flows. The conformal dimension $\Delta$ of $\mathcal{O}$ is related to the scalar mass $m$ through the AdS/CFT dictionary \cite{Witten:1998qj,Aharony:1999ti} as
\begin{equation}
(m\ell)^2 = \Delta(\Delta-d),
\end{equation}
so the deformation is only relevant ($\Delta < d$) when $m^2 < 0$. However, for each $m^2$ strictly above the Breitenlohner--Freedman bound $m^2 > -\frac{d^2}{4\ell^2}$ \cite{Breitenlohner:1982bm}, there are two possible values of $\Delta$,
\begin{equation}
\Delta_{\pm} = \frac{d}{2}\left(1 \pm \sqrt{1 + \frac{4\ell^2}{d^2}m^2}\right) \implies 0 < \Delta_- < \frac{d}{2} < \Delta_+ < d.
\end{equation}
Each $\Delta$ corresponds to specific boundary conditions on the scalar field \cite{Klebanov:1999tb}. In this paper, we will consider only $\Delta = \Delta_+ > \frac{d}{2}$. For the radial ansatz $\Phi = \phi(r)$, we can extract the source $\phi_0$ from the near-boundary profile of $\phi(r)$ \cite{DHoker:2002nbb}:\footnote{This power-law behavior of the scalar field may be obtained by taking the $r \to 0$ limit of \eqref{eomround1}, noting that $\chi(0) = 0$ and $F(0) = 1$ for asymptotically AdS geometries. This yields a simple second-order ODE.}
\begin{equation}
\phi(r) = r^{d-\Delta}\left(\phi_0 + \cdots\right) + r^{\Delta}\left(\frac{\expval{\mathcal{O}}}{2\Delta-d} + \cdots\right) \implies \phi_0 = \lim_{r \to 0} \left(\frac{1}{r}\right)^{d-\Delta}\phi(r).\label{phibdryexp}
\end{equation}
Thus, the source is determined by the leading-order term in the near-boundary expansion (the Dirichlet condition). This contrasts with the choice $\Delta = \Delta_-$ for which the source term is next-to-leading-order\footnote{To be more precise, the source term is still the coefficient of the $r^{d-\Delta}$ term, but this term becomes subleading to $r^\Delta$ near the boundary when $\Delta = \Delta_-$. This is because $\Delta_- < \frac{d}{2}$.} and is extracted by taking a derivative (the Neumann condition). Additionally, note the length dimension of $\phi_0$ is $\Delta-d$, since the scalar field itself must be dimensionless in \eqref{actscalar}.

So, $\phi_0$ is a scale that inputs into the UV data, while $\Delta$ is an additional dimensionless parameter. Furthermore, we consider the UV fixed point to be a CFT at finite temperature $\beta^{-1} = T_{\text{h}}$ and on a spatial sphere of radius $\ell$. Thus, there are three dimensionless parameters:
\begin{equation}
\text{UV data:}\ \ \left\{\frac{\beta}{\ell},\, \phi_0 \ell^{d-\Delta},\,\Delta\right\}.\label{uvData}
\end{equation}
As an aside, recall that flat topology arises in the large-volume limit of a round black hole. Specifically, we take $\beta \ll \ell$ with $\beta$ and $\ell$ kept finite, so the flows corresponding to flat black holes do not have $\frac{\beta}{\ell}$ as UV data. The only finite dimensionless parameters left are $\phi_0 \beta^{d-\Delta}$ and $\Delta$. Round black holes thus accommodate a larger parameter space of UV data.

Lastly, note that the other metric functions $\chi$ and $F$ may also be expanded around $r = 0$. Their behavior in this regime is directly determined by that of $\phi$ through the equations of motion. Specifically, we can plug \eqref{phibdryexp} into \eqref{eomround1}--\eqref{eomround3} to write
\begin{align}
\chi(r) =\ &r^{2(d-\Delta)} \left[\frac{d-\Delta}{2(d-1)}\phi_0^2 + \cdots\right] + r^d \left[\frac{2\Delta (d-\Delta)}{d(d-1)(2\Delta-d)}\phi_0\expval{\mathcal{O}} + \cdots\right]\nonumber\\
&+ r^{2\Delta} \left[\frac{\Delta \expval{\mathcal{O}}^2}{2(d-1)(2\Delta-d)^2} + \cdots\right],\label{asympchi}\\
F(r) =\ & 1 + \frac{(d-2)}{(d-\Delta)\ell^2} r^2 + \frac{\Delta \phi_0^2}{2(d-1)} r^{2(d-\Delta)} + \cdots .\label{asympF}
\end{align}
As $2(d-\Delta) < d < 2\Delta$, the leading-order term in $\chi$ is proportional to $r^{2(d-\Delta)}$. To find the first two subleading terms in $F$, we only plug-in the lowest-order terms in $\phi$ ($\sim r^{d-\Delta}$) and $\chi$ ($\sim r^{2(d-\Delta)}$).

\subsection{Trans-IR data}\label{sec:dataTIR}

Now, let us briefly comment on the near-singularity structure of black holes in the presence of matter so as to pin down what sort of ``data" would characterize the trans-IR endpoint. We again focus on scalar deformations \eqref{actscalar}, so characterizing the trans-IR data amounts to understanding free-scalar-induced backreaction on an AdS black-hole interior.

The classical evolution of the black-hole interior (or singular geometries) in the presence of matter is a deep topic dating back decades to seminal work by Belinskii, Khalatnikov, and Lifshitz (BKL) \cite{Lifshitz:1963ps,Belinskii:1970ew,Belinskii:1982pk} and by Misner \cite{Misner:1969hg}, with subsequent rigorous treatments of the dynamics utilizing the ``cosmological billiards" approach \cite{Damour:2002et,Damour:2002tc}. These methods are good near the singularity. In more recent years, the focus has shifted to numerical construction of the near-horizon region \cite{Hartnoll:2020fhc} and even to analytic study of the full interior \cite{Hartnoll:2022snh,Hartnoll:2022rdv}. Nonetheless, these different approaches are complementary to one another \cite{Henneaux:2022ijt}.

In these studies, the general structure of the black-hole interior is a Kasner universe \cite{Kasner:1921zz}. These geometries are anisotropic spacetimes. The planar version takes the form
\begin{equation}
ds_{\text{Kasner}}^2 = -\ell^2 d\tau^2 + \sum_{i=1}^{d} \tau^{2p_i} dx_i^2.
\end{equation}
The $p_i$ are called \textit{Kasner exponents}. We use these as trans-IR data. For a solution to the vacuum Einstein equations, the Kasner exponents simultaneously satisfy the constraints
\begin{equation}
\sum_{i=1}^d p_i = 1,\ \ \sum_{i=1}^d p_i^2 = 1.\label{vacKasner}
\end{equation}
Let us be more specific to our particular class of flows. Scalar fields blow up logarithmically when near a spherically-symmetric Schwarzschild singularity \cite{Doroshkevich:1978aq,Fournodavlos:2018lrk}, so we start with
\begin{equation}
\phi(r) \sim c(d-1)\log\left(\frac{r}{\ell}\right),\ \ r \gg \ell,
\end{equation}
where $c$ is some constant. We may then use the equations of motion \eqref{eomround1} and \eqref{eomround3} to write the other metric functions in this $r \gg \ell$ regime:
\begin{equation}
\chi(r) \sim c^2 (d-1)\log\left(\frac{r}{\ell}\right) + \chi_{\infty},\ \ F(r) \sim -F_\infty \left(\frac{r}{\ell}\right)^{q},
\end{equation}
where $\chi_{\infty}$ and $F_\infty > 0$ are integration constants and $q$ is shorthand for
\begin{equation}
q = d + c^2 \left(\frac{d-1}{2}\right).
\end{equation}
Plugging these into our metric ansatz \eqref{eq:metric2}, we may apply the coordinate transformation $r \to \ell \tau^{-2/q}$ to get (up to rescalings of the $t$ and $\tau$ coordinates)
\begin{equation}
ds^2 \sim -\ell^2 d\tau^2 + \tau^{2p_t} dt^2 + \left(\frac{\ell q \sqrt{F_{\infty}}}{2}\right)^2 \tau^{2p_\Omega}d\Omega_{d-1}^2.
\end{equation}
This is a ``round" Kasner universe---see \cite{Shaghoulian:2016umj}. The Kasner exponents are
\begin{equation}
p_t = 1 - \frac{2(d-1)}{q},\ \ p_\Omega = \frac{2}{q}.
\end{equation}
At this stage, it is useful to write $\phi$ as a function of $\tau$. This allows us to identify yet another Kasner exponent $p_\phi$:
\begin{equation}
\phi(\tau) = -\sqrt{2} \log \tau^{p_\phi},\ \ p_\phi = \frac{2\sqrt{(d-1)(q-d)}}{q}.
\end{equation}
Thus, we have three distinct exponents which manifestly satisfy two constraints,
\begin{equation}
p_t + (d-1)p_\Omega = 1,\ \ p_t^2 + (d-1)p_\Omega^2 + p_\phi^2 = 1,
\end{equation}
so the trans-IR data is described by just one Kasner exponent, which we choose to be $p_t$:
\begin{equation}
\text{trans-IR data:}\ \ \{p_t\}.
\end{equation}
We go from three parameters in the UV data \eqref{uvData} to one parameter, so the lossy nature of RG flow is manifest.

As an application, consider the vacuum solution for which $\phi = 0$. Then we recover the same constraint as \eqref{vacKasner}. In this case, we may exactly solve for the Kasner exponents:
\begin{equation}
\left.p_t\right|_{\text{vac}} = -1 + \frac{2}{d},\ \ \left.p_\Omega\right|_{\text{vac}} = \frac{2}{d}.\label{kasnerVacuum}
\end{equation}
As a sanity check, observe that this is consistent with the vacuum solution \eqref{vacMet}, which has a purely $d$-dependent Kasner structure that is fixed independently of UV data.

Lastly, we comment on the qualitative meaning of $p_t$ to the structure of the black-hole geometry. It (along with the other Kasner exponents) conveys information about the stability of the interior near the singularity. Specifically, the singularity is located at $\tau \to 0$, so
\begin{equation}
p_t > 0 \implies \lim_{\tau \to 0} g_{tt} \to 0,\ \ \ \ p_t < 0 \implies \lim_{\tau \to 0} g_{tt} \to +\infty.\label{gttkasner}
\end{equation}
In the former case, $g_{tt}$ decays exponentially. This decay as we approach the singularity is viewed as the ``collapse" of the Einstein--Rosen bridge \cite{Hartnoll:2020fhc,Hartnoll:2020rwq}. Meanwhile, in the latter case, we would say that the interior geometry grows near the singularity.

\subsection{From UV temperature to trans-IR exponent}

A preliminary consistency test of our intuition is to compute trans-IR data and demonstrate its functional dependence on UV data in the presence of a relevant deformation. This should contrast with the vacuum case, for which the Kasner exponents are fixed at \eqref{kasnerVacuum}.

To do this, we first numerically construct a large array of black-hole solutions to the equations of motion \eqref{eomround1}--\eqref{eomround3}. Doing so requires fixing the AdS radius; we set $\ell = 1$. The finer specifics of our construction procedure are discussed in more detail in Appendix \ref{app:scalarNums}, but at this stage we note that $d$ and $\Delta$ are fixed. With the solutions in hand, we then select for metrics with specific preselected values of the deformation parameter $\phi_0$ up to some small error, which we take to be $0.5\%$ throughout our numerics (or $0.005$ for $\phi_0 = 0$). This is done prior to subsequent calculations to save on computational resources.

\begin{figure}
\centering
\includegraphics[scale=0.7]{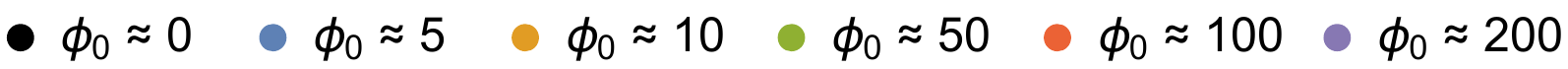}
\vspace{-0.5cm}\\
\subfloat[$d = 2$, $\Delta = \frac{3}{2}$]{
\includegraphics[scale=0.55]{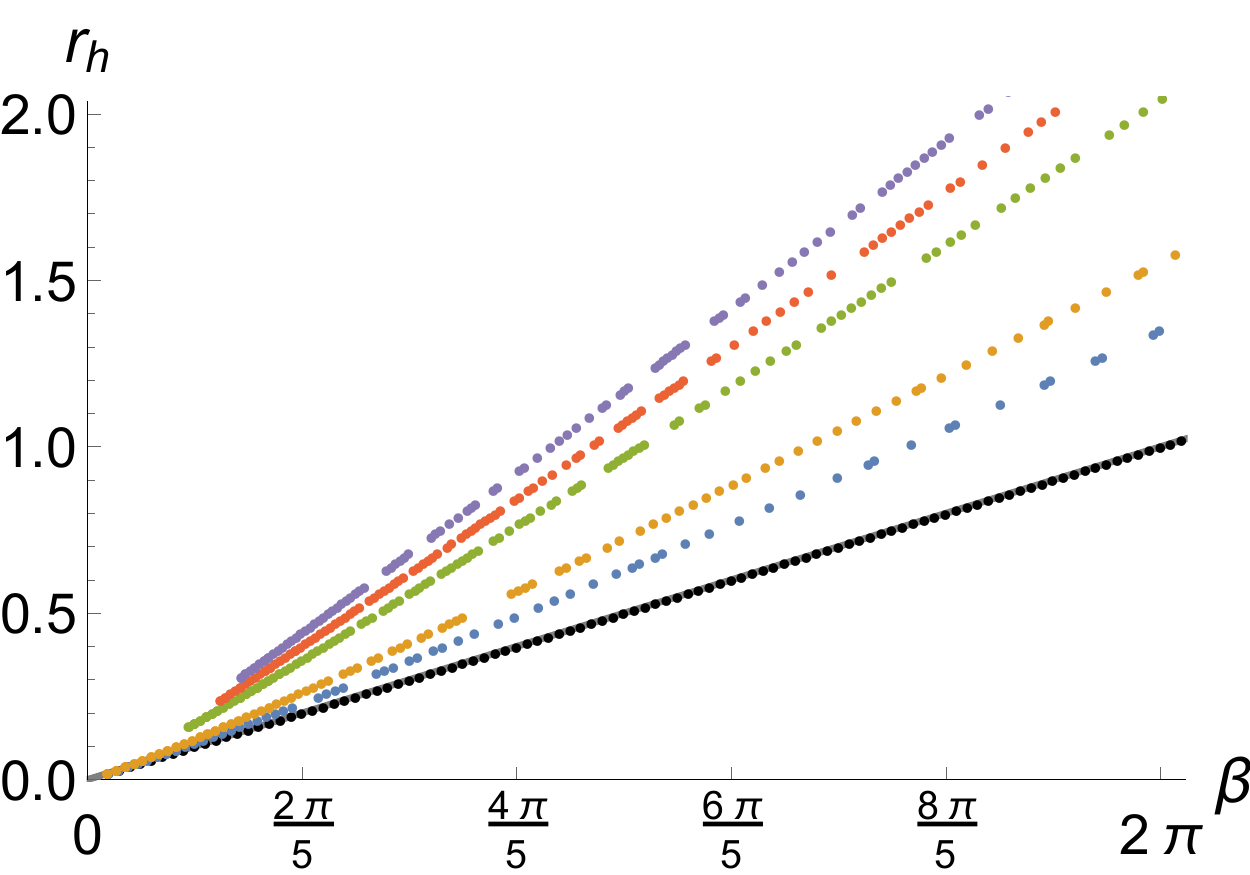}
}\ \ \ \ 
\subfloat[$d = 3$, $\Delta = 2$]{
\includegraphics[scale=0.55]{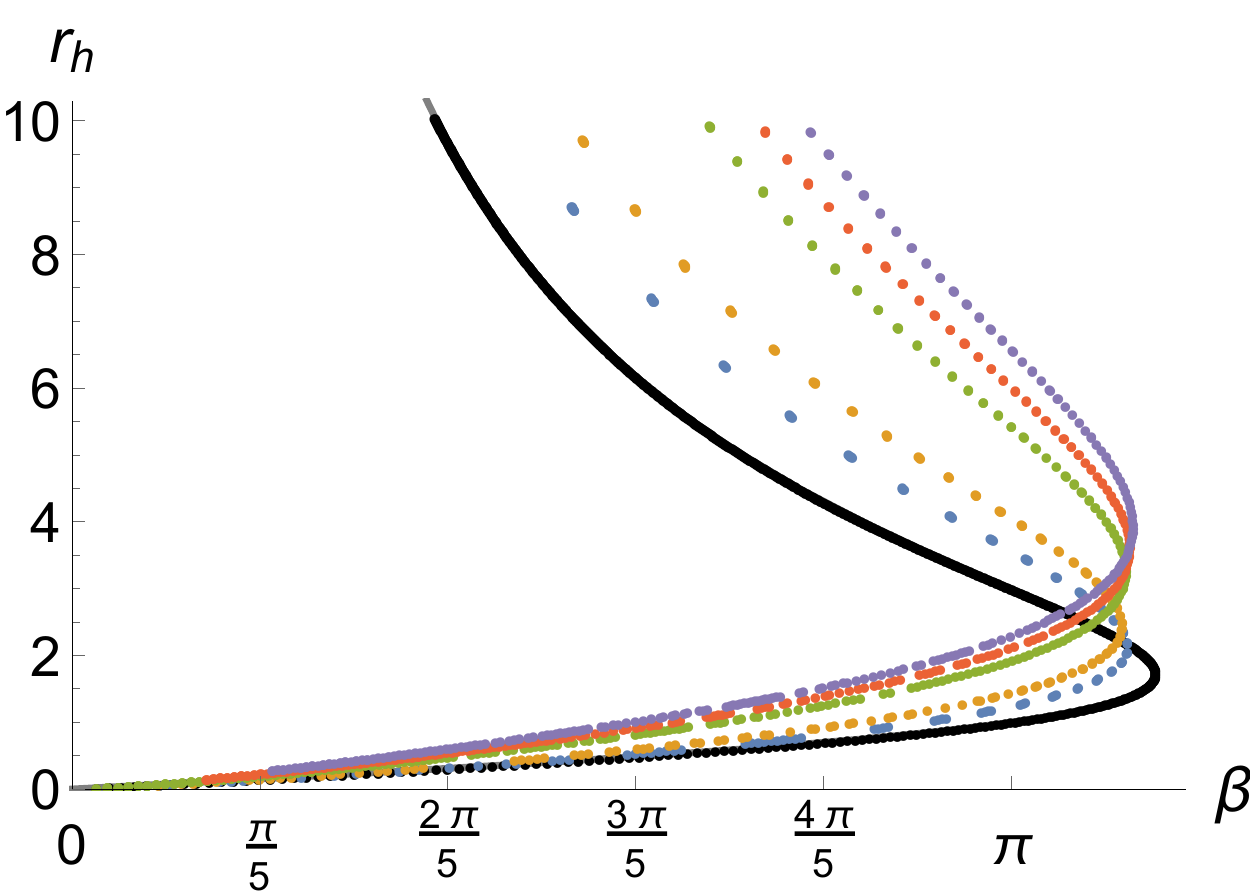}
}
\caption{Plots of the horizon radius $r_{\text{h}}$ versus inverse temperature $\beta$ with $\ell = 1$ and for both (a) $d = 2$, $\Delta = \frac{3}{2}$ and (b) $d = 3$, $\Delta = 2$. We plot these points for various different values of the deformation parameter $\phi_0 \pm 0.5\%$. For $d = 2$, we observe a one-to-one linear relationship, indicating that each temperature corresponds to just a single black hole in the bulk. For $d = 3$ however, there are two solutions for each $\beta$---a large black hole (the lower branches) and a small black hole (the upper branches). As a sanity check, we note the $\phi_0 \approx 0$ numerical points are consistent with known analytic expressions for $r_{\text{h}}$, which are $\frac{\beta}{2\pi}$ for $d = 2$ and $\frac{2\pi}{\beta} \pm \sqrt{\left(\frac{2\pi}{\beta}\right)^2 - 3}$ (resp. the small and large branches) for $d = 3$.}
\label{figs:largesmallBHs}
\end{figure}

For $d = 2$, each temperature only furnishes a single black-hole geometry---a deformed static BTZ black hole \cite{Banados:1992wn,Banados:1992gq}. However, it is well known that for $d > 2$ bulk spatial dimensions there are two branches of black hole solutions (in terms of the horizon radius $r_{\text{h}}$) for each $\beta$---large black holes and small black holes. This is not only true in vacuum but also with a deformation turned on. We illustrate this point in Figure \ref{figs:largesmallBHs}. As a bonus, the $d = 3$ plot also exemplifies how $d > 2$ theories include a maximal $\beta$ or minimal temperature (the ``spinodal" point) beyond which there are no black hole solutions.

In our coordinates \eqref{eq:metric2}, the conformal boundary is at $r = 0$. Thus if there are two solutions at some temperature, the large black hole corresponds to the smaller value of $r_{\text{h}}$ while the small black hole corresponds to the large value. Although there is this ambiguity in the bulk geometry, we emphasize that the small black hole is thermodynamically unstable. Nonetheless, we may view it as some subdominant (in the canonical ensemble) thermal state with its own trans-IR flow.

\begin{figure}
\centering
\includegraphics[scale=0.7]{figs/legendDeformations.pdf}
\vspace{-0.5cm}\\
\subfloat[$d = 2$, $\Delta = \frac{3}{2}$\label{figs/ptvsbeta-2d}]{
\includegraphics[scale=0.55]{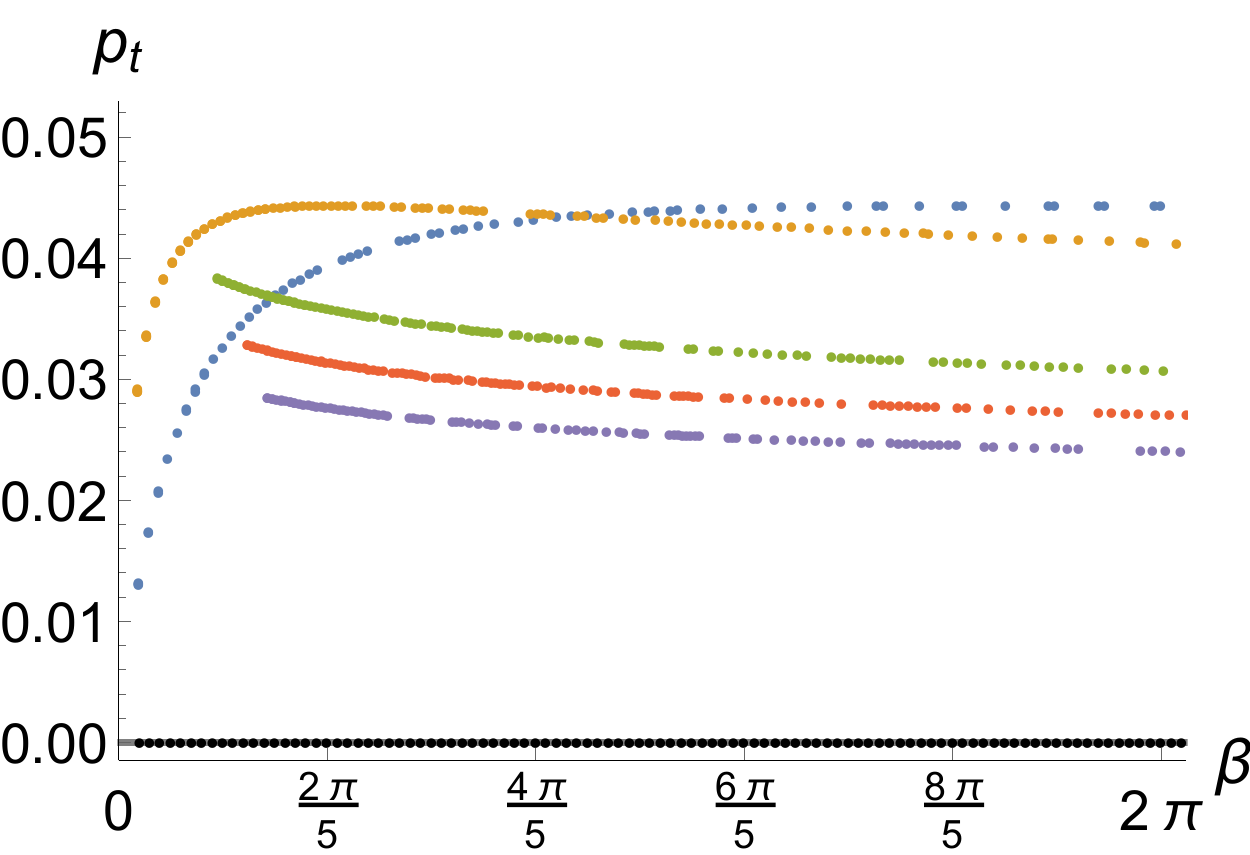}
}\ \ \ \ 
\subfloat[$d = 3$, $\Delta = 2$\label{figs:plotsptbetad3}]{
\includegraphics[scale=0.55]{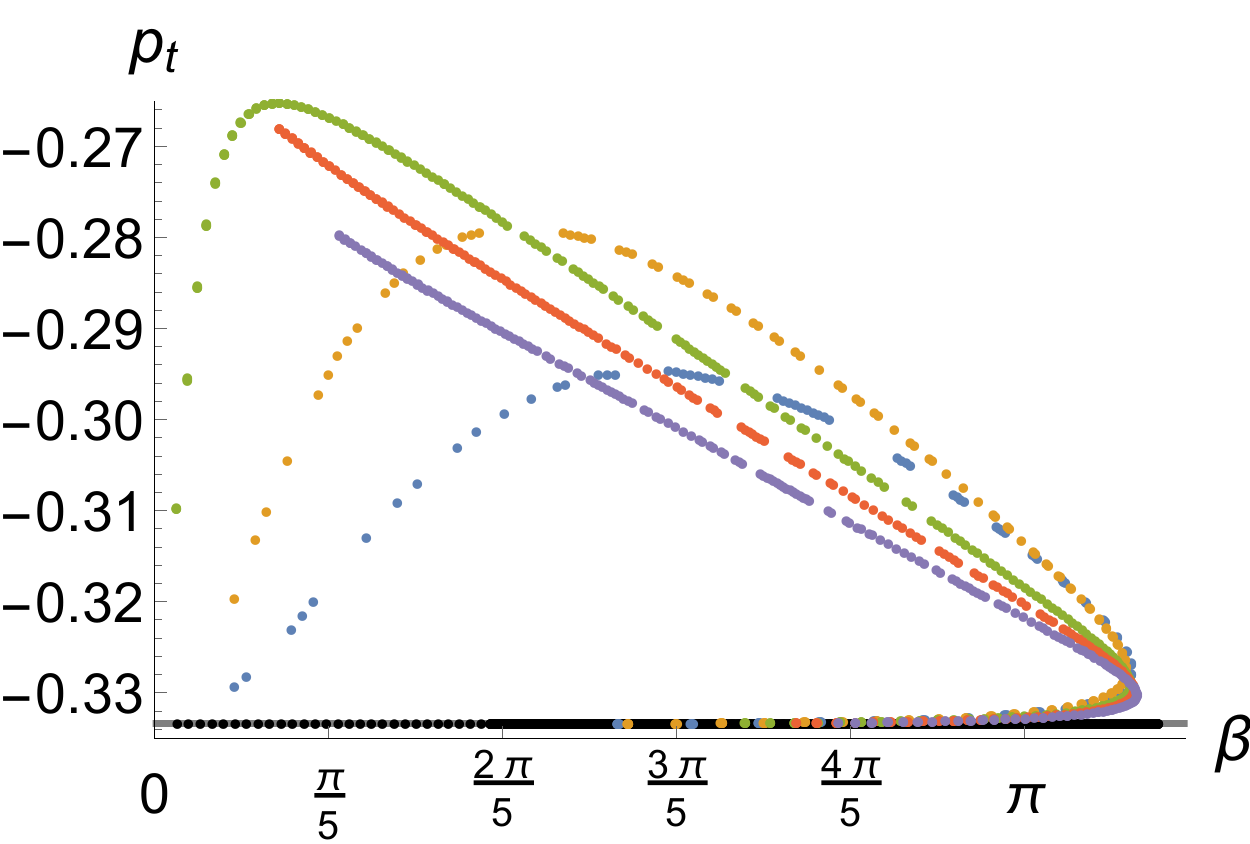}
}
\caption{Plots of the Kasner exponent $p_t$ versus inverse temperature $\beta$ with $\ell = 1$ and for both (a) $d = 2$, $\Delta = \frac{3}{2}$ and (b) $d = 3$, $\Delta = 2$. We again plot these points for various different values of the deformation parameter $\phi_0 \pm 0.5\%$. For both, the $\phi_0 \approx 0$ values of $p_t$ are only fixed by the dimension \eqref{kasnerVacuum} and thus respectively are $0$ and $-\frac{1}{3}$. For $\phi_0 \not\approx 0$, the $d = 2$ plot is a well-defined function whereas the $d = 3$ plot is branched. In the latter, the upper branch corresponds to large black holes while the lower branch corresponds to small black holes.}
\label{figs:ptbetaplots}
\end{figure}

With that in mind, for each of these families of black holes we also plot $p_t$ as a function of the inverse temperature in Figure \ref{figs:ptbetaplots}. For both $d = 2$ and $d = 3$, we see that the vacuum Kasner exponent $p_t|_{\text{vac}} = -1 + \frac{2}{d}$ \eqref{kasnerVacuum} are reproduced by our numerics. With a deformation turned on however, $p_t$ develops a nontrivial relationship with $\beta$.

For $d = 2$, we get a fairly consistent functional structure (at least for low $\phi_0$, but we expect it to hold for larger $\phi_0$). $p_t$ apparently increases from the vacuum value of $0$ for small $\beta$ until reaching a maximum value dependent on $\phi_0$, and then decreases. We may even posit that the value of $\beta$ at this maximum decreases with $\phi_0$.

Meanwhile for $d = 3$, the presence of two black holes for each $\beta$ manifests branching in Figure \ref{figs:plotsptbetad3}. More specifically, we observe that the small black holes have Kasner exponents much closer to the vacuum value than than those of the large black holes. In other words, the small black holes are the lower branch of Figure \ref{figs:plotsptbetad3}, while the large black holes exhibit a similar behavior to those of $d = 2$ with a spinodal point present. One way to interpret this is to say that small black holes have ``more" near-singularity geometry, since as per \eqref{gttkasner} $g_{tt}$ would blow-up more for small black holes.

So to summarize, deformations of the UV turn on nontrivial relationships between UV and trans-IR data. In particular, the near-singularity geometry is no longer completely specified by $d$. We now want to describe such relationships in terms of other parameters of the UV CFT that are directly connected to phase transitions of holographic quantities.

\section{Imprints of entanglement plateaux} \label{sec:plateaux}

One phase transition unique to the round black hole concerns holographic entanglement entropy as computed by the Ryu--Takayanagi (RT) prescription \cite{Ryu:2006bv,Hubeny:2007xt}, which we now summarize. First, consider a bulk codimension-2 surface $\Gamma$ which is ``homologous" to $\mathcal{R}$ ($\Gamma \sim \mathcal{R}$), by which we mean that there exists a codimension-1 bulk region $\Sigma$ for which
\begin{equation}
\partial\Sigma = \mathcal{R} \cup \Gamma,\ \ \partial\mathcal{R} = \partial\Gamma,\label{homologyCondition}
\end{equation}
Then, to leading order in a small (in units of AdS radius) $G_{\text{N}}$ expansion, the entanglement entropy of a boundary interval $\mathcal{R}$ on a fixed-time slice is proportional to the area of the smallest such $\Gamma$:
\begin{equation}
S(\mathcal{R}) =\, \stackrel[\Gamma \sim \mathcal{R}]{}{\text{min}\,\text{ext}}\left[\frac{\text{Area}(\Gamma)}{4G_{\text{N}}}\right].
\end{equation}
The minimal-area extremal surface is called the RT surface.

In simple cases like pure AdS or the planar AdS-Schwarzschild black hole, the RT surface is connected if $\mathcal{R}$ is connected. However, this need not be the case in the round black hole (as also appreciated in other work \cite{Headrick:2007km,Azeyanagi:2007bj,Blanco:2013joa}). For example, when $\mathcal{R}$ constitutes a sufficiently large connected subregion of the boundary in a higher-dimensional ($d > 2$) round AdS black hole, one does not even have connected homologous extremal surfaces as candidates for the RT surface in the first place \cite{Hubeny:2013gta}! The relevant feature of a round black hole is that any bulk region $\Sigma$ bounded by a sufficiently large $\mathcal{R}$ and some extremal connected surface always includes the horizon, so it must be included as a separate connected component of the RT surface.\footnote{Note that the horizon is a topologically closed surface and thus has an empty boundary, so it can be included in $\Gamma$ without breaking the homology condition \eqref{homologyCondition}.} Indeed, allowing for disconnected $\Gamma$ is the only way to have the RT prescription be continuous with the Bekenstein--Hawking formula  \cite{Bekenstein:1973ur,Hawking:1976de} in the limit where $\mathcal{R}$ is the full boundary, for which the RT surface should \textit{only} be the horizon.

As a generic adaptation of the RT prescription to round black holes, \cite{Hubeny:2013gta} starts by considering two separate classes of codimension-2 surfaces homologous to $\mathcal{R}$---those which are connected and those which are disconnected and include the horizon. See Figure \ref{figs:roundEntanglement} for a visual representation. Then, computing the entanglement entropy amounts to finding the minimum-area extremal surface\footnote{If one of these classes does not include any extrema (like the $d > 2$ round AdS black holes \cite{Hubeny:2013gta}), then extrema must exist solely the other class.} among both classes:
\begin{equation}
S(\mathcal{R}) = \text{min} \left[\frac{\text{Area}(\Gamma_{\text{C}})}{4G_{\text{N}}},\frac{\text{Area}(\Gamma_{\text{D}})}{4G_{\text{N}}}\right] = \text{min} \left(S_{\text{C}},\,\overline{S}_{\text{D}} + S_{\text{BH}}\right).\label{entropiesRCD}
\end{equation}

\input{figs/roundEntanglement}

\noindent In the equation above, $\Gamma_{\text{C}}$ denotes a generic extremal surface which is both connected and homologous to $\mathcal{R}$. The area of this surface is directly related to $S_{\text{C}}$. Meanwhile $\Gamma_{\text{D}}$ is the extremal surface which is disconnected and homologous to $\mathcal{R}$. It helps to break the resulting entropy into two terms, each corresponding to a connected component of $\Gamma_{\text{D}}$. The first term $\overline{S}_{\text{D}}$ is computed by the connected component $\overline{\Gamma}_{\text{D}} \subset \Gamma_{\text{D}}$ which is anchored to the conformal boundary and wraps around the horizon. The second term $S_{\text{BH}}$ is the Bekenstein--Hawking entropy of the horizon itself.

That there are two competing candidates in the round black hole allows for a first-order phase transition as described in \cite{Hubeny:2013gta}. First, recall that if the full system is in a pure state, then the entanglement entropy of any subregion $\mathcal{R}$ and its complement $\mathcal{R}^c$ must match. For mixed states, we may quantify the ``deviation" from this purity as
\begin{equation}
\delta S_{\mathcal{R}} = S(\mathcal{R}) - S(\mathcal{R}^c).
\end{equation}
By the Araki--Lieb inequality (cf. Theorem 2c of \cite{Araki:1970ba}),
\begin{equation}
|\delta S_{\mathcal{R}}| \leq S(\mathcal{R} \cup \mathcal{R}^c) = S_{\text{BH}},\label{alineq}
\end{equation}
where the entropy of the full boundary system is equated to the Bekenstein--Hawking entropy. Saturation of the Araki--Lieb inequality corresponds to a canonical factorization of the full boundary degrees of freedom, with any of the resulting factors carrying all of the microscopic entanglement entropy (cf. Theorem III.2 of \cite{Zhang_2011}).

We may analyze the deviation from purity on the left-hand side of \eqref{alineq} as we tune the ``size" (we make this more precise below) of $\mathcal{R}$ from $0$ to half of the full boundary interval.\footnote{Because we are comparing the entropy of $\mathcal{R}$ against that of its complement, taking $\mathcal{R}$ to be half of the boundary is the same as considering the extremal case $\mathcal{R} = \mathcal{R}^c$. Thus by the $\mathbb{Z}_2$ symmetry which exchanges the roles of $\mathcal{R}$ and $\mathcal{R}^c$, considering intervals $\mathcal{R}$ that are more than half of the full boundary is redundant.} In doing so, we find that for small $\mathcal{R}$ there is a window of interval sizes for which $S_{\mathcal{R}}$ and $S_{\mathcal{R}^c}$ are respectively computed by $\Gamma_{\text{C}}$ and $\Gamma_{\text{C}} \cup \Gamma_{\text{h}}$, where $\Gamma_{\text{h}}$ is the horizon. For $\mathcal{R}$ near half of the boundary, however, there is another window of interval sizes for which \textit{both} entropies are computed by corresponding connected phases ($\Gamma_C$ and $\overline{\Gamma}_D$, respectively). According to \eqref{entropiesRCD}, these two windows are respectively described by
\begin{align}
S_{\text{C}} \leq \overline{S}_{\text{D}} - S_{\text{BH}} &\implies |\delta S_{\mathcal{R}}| = S_{\text{BH}},\label{windowSmall}\\
\overline{S}_{\text{D}} - S_{\text{BH}} < S_{\text{C}} \leq \overline{S}_{\text{D}} &\implies 0 \leq |\delta S_{\mathcal{R}}| < S_{\text{BH}}.\label{windowLarge}
\end{align}
If we plot $|\delta S_{\mathcal{R}}|$ as a function of interval size, we observe a so-called entanglement ``plateau" \cite{Hubeny:2013gta}. The fall from the plateau corresponds to loss of saturation of the Araki--Lieb inequality.

We aim to understand how this phase transition imprints upon the trans-IR Kasner exponents. To do so, we take $\mathcal{R}$ to be a round ``cap" [with $SO(d-1)$ symmetry] in the boundary whose size is controlled by a single angular parameter $\theta_{\mathcal{R}}$. The transition is found to occur at a particular $\theta_{\mathcal{R}} = \theta_{\text{k}}$. With a scalar deformation in the UV, we find that this transition point can be viewed as UV data in lieu of the CFT temperature parameter $\frac{\beta}{\ell}$ in the list \eqref{uvData}
. With that in mind, we plot $\theta_{\text{k}}$ against the Kasner exponent $p_t$ at fixed values of the deformation parameter $\phi_0 \ell^{d-\Delta}$.

As a matter of practicality, we focus on the $d = 2$ case for which the extremal surfaces can be computed from a first-order ODE. The higher-dimensional cases are in principle possible to analyze, but they require numerically solving second-order ODEs. We discuss the basic machinery of general $d \geq 2$ in Appendix \ref{app:entropyCaps}, leaving the higher-dimensional problem open for future work.

Essentially, we will find that a finite, nonzero deformation induces a nontrivial relationship between $p_t$ and $\theta_{\text{k}}$. The takeaway is that a deformation allows entanglement structure in the UV to imprint upon the trans-IR data (at least in the low-dimensional case of $d = 2$). We find that this effect is numerically small.

\subsection{Ryu--Takayanagi in deformations of the BTZ black hole} 

We compute entanglement plateaux for deformations of the round BTZ black hole in $d = 2$, which is described by the metric (fixing the curvature radius as $\ell = 1$)
\begin{equation}
ds^2 = \frac{1}{r^2}\left[-e^{-\chi(r)} F(r) dt^2 + \frac{dr^2}{F(r)} + d\theta^2\right],
\end{equation}
with $t \in \mathbb{R}$, $r > 0$, and $\theta \sim \theta + 2\pi$. Recall that we assume $F$ to have a simple root at $r = r_{\text{h}}$, defined as the black hole horizon. Furthermore, it is convenient to take the fundamental domain $\theta \in (-\pi,\pi]$, with $\mathcal{R}$ at fixed time being parameterized as $\theta \in [-\theta_{\mathcal{R}},\theta_{\mathcal{R}}]$ for some $0<\theta_{\mathcal{R}}<\frac{\pi}{2}$.\footnote{Recall that we are only considering $\mathcal{R}$ up to half of the full boundary.} Our goal will be to find the value $\theta_{\mathcal{R}} = \theta_{\text{k}}$ at which the Araki--Lieb inequality switches between being saturated and holding strictly.

To compute the extremal surfaces, we first use the metric to write the area functional of a surface $\theta = \theta(r)$ at fixed $t$ (omitting bounds of integration for now)
\begin{equation}
\mathcal{A} = \int \frac{dr}{r} \sqrt{\frac{1}{F(r)} + \theta'(r)^2}.\label{areaFunc}
\end{equation}
We then use the Euler-Lagrange equation to obtain the equation of motion for the extremal surfaces,
\begin{equation}
\theta'(r)^2 = \frac{r^2}{F(r) (r_*^2 - r^2)},\label{eom2d}
\end{equation}
Here, $r_* < r_{\text{h}}$ is a bulk constant of motion characterizing the turnaround point of the extremal surface, i.e., $\frac{dr}{d\theta}|_{r = r_*} = \frac{1}{\theta'(r_*)} = 0$ and $r \in [0,r_*]$. By integrating \eqref{eom2d}, we can relate this $r_*$ parameter to $\theta_{\mathcal{R}}$.

We now have the necessary expressions to compute the areas of the RT candidates for both $\mathcal{R}$ and $\mathcal{R}^c$. Before we proceed, we reiterate that $\Gamma_{\text{C}}$ denotes the connected extremal surface homologous to $\mathcal{R}$ while the homologous disconnected extremal surface $\Gamma_{\text{D}}$ decomposes into connected components $\overline{\Gamma}_{\text{D}} \cup \Gamma_{\text{h}}$, where $\Gamma_{\text{h}}$ is the horizon and $\partial\overline{\Gamma}_{\text{D}} = \partial\mathcal{R}$. We observe that the connected phase for the entropy of the complement $\mathcal{R}^c$ is then computed by the connected component $\overline{\Gamma}_{\text{D}}$, and the disconnected phase is computed by the union $\Gamma_{\text{C}} \cup \Gamma_{\text{h}}$. So at this stage, we simply need to compute the endpoints and areas of $\Gamma_{\text{C}}$ and $\overline{\Gamma}_{\text{D}}$ in terms of their turnaround points.

\paragraph{Computing $\Gamma_{\text{C}}$}

We first relate the turnaround point $r_{\text{c}*}$ of $\Gamma_{\text{C}}$ to $\theta_{\mathcal{R}}$. To start, note that $\Gamma_{\text{C}}$ represents a surface which does not wrap around the black hole horizon relative to $\mathcal{R}$ (Figure \ref{figs:roundEntanglement}). Therefore, the $\theta(r) > 0$ branch has the negative root of \eqref{eom2d} as its derivative, so we write
\begin{equation}
\theta(r_{\text{c}*}) - \theta(0) = -\int_0^{r_{\text{c}*}} \frac{dr}{\sqrt{F(r)}}\frac{r}{\sqrt{r_{\text{c}*}^2 - r^2}}.
\end{equation}
By the symmetry of our parameterization, $\theta(r_{\text{c}*}) = 0$. Additionally, $\theta(0) = \theta_{\mathcal{R}}$. As such, we have
\begin{equation}
\label{eq:rc}
\theta_{\mathcal{R}} = \int_{0}^{r_{\text{c}*}} \frac{dr}{\sqrt{F(r)}}\frac{r}{\sqrt{r_{\text{c}*}^2 - r^2}}.
\end{equation}
Furthermore, the area of $\Gamma_{\text{C}}$ may be written by plugging \eqref{eom2d} directly into \eqref{areaFunc}:
\begin{equation}
\text{Area}(\Gamma_{\text{C}}) = 2\int_{0}^{r_{\text{c}*}} \frac{dr}{\sqrt{F(r)}} \frac{r_{\text{c}*}/r}{\sqrt{r_{\text{c}*}^2 - r^2}}.\label{areaCbh}
\end{equation}

\paragraph{Computing $\overline{\Gamma}_{\text{D}}$}

We now relate the turnaround point $r_{\text{d}*}$ of $\overline{\Gamma}_{\text{D}}$ to $\theta_{\mathcal{R}}$. As shown in Figure \ref{figs:roundEntanglement}, this surface wraps around the black hole horizon relative to $\mathcal{R}$. As such, the derivative of the $\theta(r) > 0$ branch is the positive root of \eqref{eom2d}, so
\begin{equation}
\theta(r_{\text{d}*})- \theta(0)  = \int^{r_{\text{d}*}}_0 \frac{dr}{\sqrt{F(r)}} \frac{r}{\sqrt{r_{\text{d}*}^2 - r^2}}.
\end{equation}
This time, we have that $\theta(r_{\text{d}*}) = \pi$ and $\theta(0) = \theta_{\mathcal{R}}$. Hence,
\begin{equation}
\label{eq:rd}
\theta_{\mathcal{R}} = \pi - \int_0^{r_{\text{d}*}} \frac{dr}{\sqrt{F(r)}} \frac{r}{\sqrt{r_{\text{d}*}^2 - r^2}}.
\end{equation}
However, the area of $\overline{\Gamma}_{\text{D}}$ takes the same form as that of $\Gamma_{\text{C}}$,
\begin{equation}
\text{Area}(\overline{\Gamma_{D}}) = 2\int_{0}^{r_{\text{d}*}} \frac{dr}{\sqrt{F(r)}} \frac{r_{\text{d}*}/r}{\sqrt{r_{\text{d}*}^2 - r^2}}.\label{areaDbh}
\end{equation}

\paragraph{Applying Ryu--Takayanagi}

For a fixed $\mathcal{R}$ (and $\mathcal{R}^c$), we can in principle use the simultaneous constraints \eqref{eq:rc} and \eqref{eq:rd} to solve for the turnaround points $r_{\text{c}*}$ and $r_{\text{d}*}$ in terms of the interval size parameter $\theta_{\mathcal{R}}$. With those in hand, we can then plug the expressions \eqref{areaCbh} and \eqref{areaDbh} into the holographic prescriptions for the entanglement entropies of $\mathcal{R}$ and $\mathcal{R}^c$ to evaluate the entropies:
\begin{align}
S(\mathcal{R}) &= \frac{1}{4G_{\text{N}}}\text{min}\left[\text{Area}(\Gamma_{\text{C}}),\,\text{Area}(\overline{\Gamma}_{\text{D}}) + \frac{2\pi}{r_{\text{h}}}\right],\\
S(\mathcal{R}^c) &= \frac{1}{4G_{\text{N}}}\text{min}\left[\text{Area}(\Gamma_{\text{C}}) + \frac{2\pi}{r_{\text{h}}},\,\text{Area}(\overline{\Gamma}_{\text{D}})\right].
\end{align}
The $\frac{2\pi}{r_{\text{h}}}$ term is the horizon area in the $d = 2$ round black hole and notably does not depend on the deformation. With these equations, we may compute the absolute difference of entropies $|\delta S_{\mathcal{R}}| = |S(\mathcal{R}) - S(\mathcal{R}^c)|$ explicitly. By scanning over $\theta_{\mathcal{R}}$, we can plot $|\delta S_{\mathcal{R}}|$ as a function of interval size, and it is in such plots that we identify entanglement plateaux.

That being said, we know that the transition in entropy occurs between the windows \eqref{windowSmall} and \eqref{windowLarge}, i.e.,
\begin{equation}
\theta_{\mathcal{R}} = \theta_{\text{k}} \iff S_{\text{C}} = \overline{S}_{\text{D}} - S_{\text{BH}} \iff \frac{2\pi}{r_{\text{h}}} = \text{Area}(\overline{\Gamma}_{\text{D}}) - \text{Area}(\Gamma_{\text{C}}).
\end{equation}
This along with \eqref{eq:rc} and \eqref{eq:rd} completely constrain the transition point and can be used to find the transition value $\theta_{\mathcal{R}} = \theta_{\text{k}}$ without needing to plot $|\delta S_{\mathcal{R}}|$.

\subsection{Plateaux kinks versus Kasner exponents}

Given some solution to the equations of motion with horizon radius $r = r_{\text{h}}$, we reiterate that the respective turnaround points $r_{\text{c}*}$ and $r_{\text{d}*}$ of $\Gamma_{\text{C}}$ and $\overline{\Gamma}_{\text{D}}$ at the transition angle $\theta_{\mathcal{R}} = \theta_{\text{k}}$ are fixed by the following system of constraints:
\begin{align}
\theta_{\text{k}} &= \int_{0}^{r_{\text{c}*}} \frac{dr}{\sqrt{F(r)}}\frac{r}{\sqrt{r_{\text{c}*}^2 - r^2}},\\
\theta_{\text{k}} &= \pi - \int_{0}^{r_{\text{d}*}} \frac{dr}{\sqrt{F(r)}}\frac{r}{\sqrt{r_{\text{d}*}^2 - r^2}},\\
\frac{\pi}{r_{\text{h}}} &= \int_{0}^{r_{\text{d}*}} \frac{dr}{\sqrt{F(r)}}\frac{r_{\text{d}*}/r}{\sqrt{r_{\text{d}*}^2 - r^2}} - \int_{0}^{r_{\text{c}*}} \frac{dr}{\sqrt{F(r)}}\frac{r_{\text{c}*}/r}{\sqrt{r_{\text{c}*}^2 - r^2}}.
\end{align}
To use these constraints, we first obtain a large number of numerical solutions to the equations of motion \eqref{eomround1}--\eqref{eomround3}. We then scan for fixed values of the deformation parameter $\phi_0$ (with some prescribed uncertainty) and, for this subset of solutions, numerically solve the above constraints. After \textit{post hoc} validation that the numerical output indeed solves these constraints, we are left with the desired values of $\theta_{\text{k}}$. We can then check whether $\theta_{\text{k}}$ is a monotonic function of $\beta$, thereby allowing us to substitute the latter for the former in the UV data without issue. We then plot the Kasner exponent $p_t$ against $\theta_{\text{k}}$.

Before discussing the numerics, however, we note that the case of the (undeformed) BTZ black hole can be understood analytically. We do so both as a warm-up and to have a consistency test in hand for our numerics.

\paragraph{Undeformed BTZ}

\begin{figure}
\centering
\includegraphics[scale=0.75]{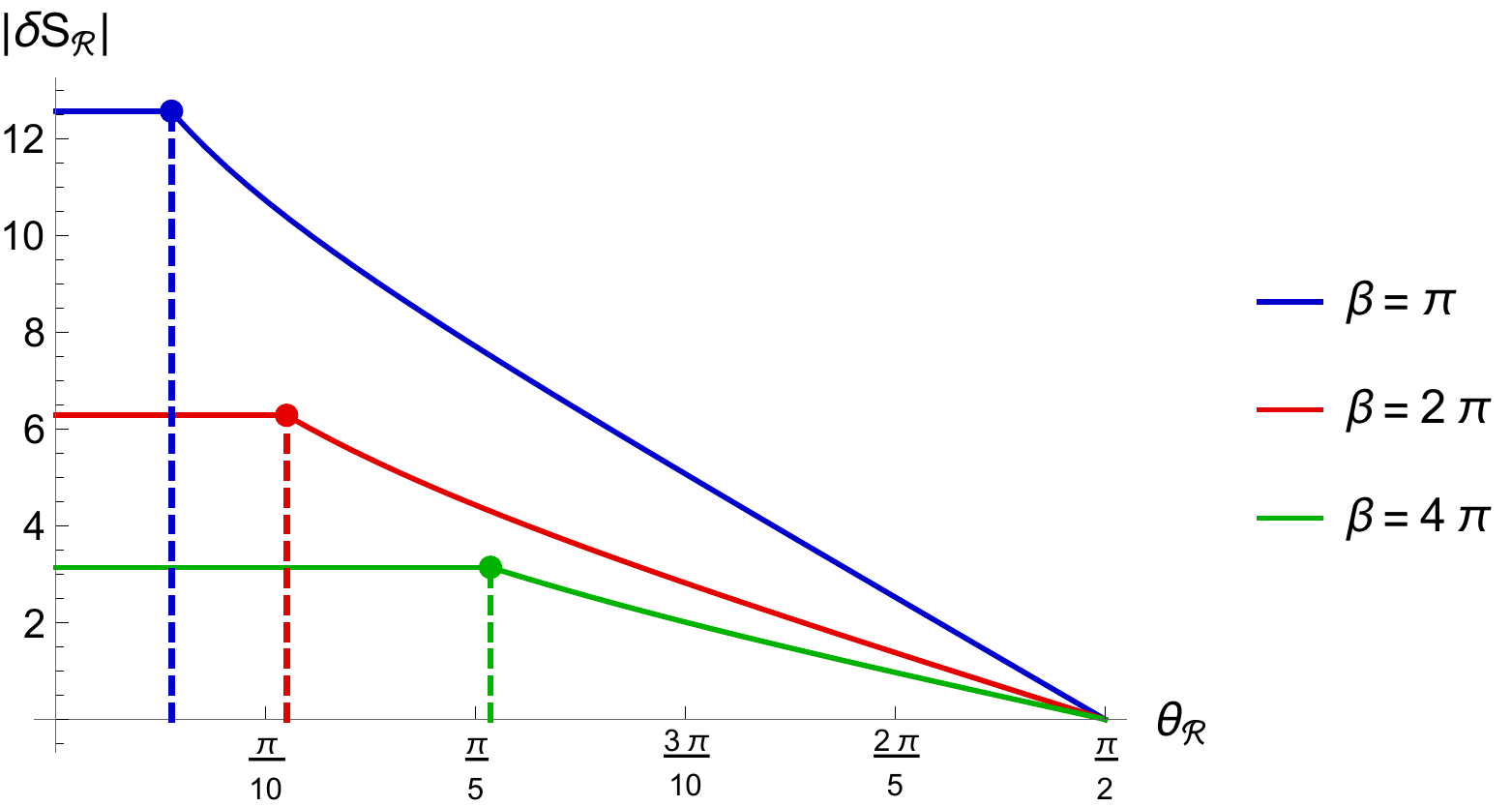}
\caption{Exact entanglement plateaux for the BTZ black hole (with $\ell = 1$) for inverse temperatures $\beta = \pi$ (blue), $\beta = 2\pi$ (red), and $\beta = 4\pi$ (green). The kinks in each plot are the points at which the phase of $\delta S_{\mathcal{R}}$ (computed via the RT prescription and with $4G_{\text{N}} = 1$) changes. These respectively occur at $\theta_{\mathcal{R}} \approx 0.055\pi$, $\theta_{\mathcal{R}} \approx 0.110\pi$, and $\theta_{\mathcal{R}} \approx 0.207\pi$.} 
\label{figs:entPlatHP}
\end{figure}

Let us review the case of the BTZ black hole in pure gravity, also analyzed by \cite{Hubeny:2013gta}. As it turns out, the interval size corresponding to the plateau of the BTZ can be computed analytically. First, recall that
\begin{equation}
F_{\text{BTZ}}(r) = 1 - \frac{r^2}{r_{\text{h}}^2}.
\end{equation}
This can be used to compute both $r_{\text{c}*}$ and $r_{\text{d}*}$ as functions of $\theta_{\mathcal{R}}$:
\begin{align}
r_{\text{c}*}^{\text{BTZ}} &= r_{\text{h}} \tanh\left(\frac{\theta_{\mathcal{R}}}{r_{\text{h}}}\right),\label{connectedr}\\
r_{\text{d}*}^{\text{BTZ}} &= r_{\text{h}} \tanh\left(\frac{\pi-\theta_{\mathcal{R}}}{r_{\text{h}}}\right).\label{disconnectedr}
\end{align}
Now, we want to compute the interval size $\theta_{\mathcal{R}} \equiv \theta_{\text{k}}^{\text{BTZ}}$ defining the kink of the plateau.\footnote{Note that we may also use the interval size at which the RT surface of $\mathcal{R}$ changes phase, which is computed by \cite{Hubeny:2013gta}. However, this is redundant. The phase transition in the surface for $\mathcal{R}$ occurs in the domain where it is more than half of the boundary system, and it precisely coincides with where the kink develops if we instead analyze the entanglement plateau of the complementary interval $\mathcal{R}^c$.} Recalling \eqref{windowSmall}--\eqref{windowLarge}, this is found by setting
\begin{equation}
\text{Area}(\overline{\Gamma}_{\text{D}}) - \text{Area}(\Gamma_{\text{C}}) = \frac{2\pi}{r_{\text{h}}}.\label{constraintKink}
\end{equation}
For the BTZ geometry, we can compute each of these areas, and thus $\theta_{\text{k}}^{\text{BTZ}}$, analytically. First, define a regulator surface $r = \epsilon$. The regulated areas are then
\begin{align}
\text{Area}(\Gamma_{\text{C}})_{\text{BTZ}}^{\text{reg}} &= -2 \log\left(\frac{\epsilon}{2r_h}\right) + \log\left(\frac{r_{\text{c}*}^2}{r_{\text{h}}^2 - r_{\text{c}*}^2}\right),\label{connected}\\
\text{Area}(\overline{\Gamma}_{\text{D}})_{\text{BTZ}}^{\text{reg}} &= -2\log\left(\frac{\epsilon}{2r_h}\right) + \log\left(\frac{r_{\text{d}*}^2}{r_{\text{h}}^2 - r_{\text{d}*}^2}\right).\label{disconnected}
\end{align}
The UV divergences manifestly cancel when we take the difference, so the constraint \eqref{constraintKink} is well-defined and yields the following relation:
\begin{equation}
\theta_{\text{k}}^{\text{BTZ}} = \frac{\pi}{2} - \frac{r_{\text{h}}}{2}\log\cosh\left(\frac{\pi}{r_{\text{h}}}\right).
\end{equation}
The horizon temperature \eqref{hawkTemp2} of the BTZ black hole is $T_{\text{h}} = \frac{1}{2\pi r_{\text{h}}}$, so we may write the kink angle as a monotonic function of the inverse CFT temperature $\beta = T_{\text{h}}^{-1}$,
\begin{equation}
\theta_{\text{k}}^{\text{BTZ}} = \frac{\pi}{2}\left[1 - \frac{\beta}{2\pi^2}\log\cosh\left(\frac{2\pi^2}{\beta}\right)\right].\label{analyticKBTZBeta}
\end{equation}
So the value of $\theta_{\text{k}}^{\text{BTZ}}$ depends on the temperature (or size through $\beta \sim r_{\text{h}}$) of the black hole. For high-temperature ($\beta \to 0$) and low-temperature ($\beta \to \infty$) BTZ black holes,
\begin{equation}
\lim_{\beta \to 0} \theta^{\text{BTZ}}_{\text{k}} = 0,\ \ \lim_{\beta \to \infty} \theta^{\text{BTZ}}_{\text{k}} = \frac{\pi}{2}.
\end{equation}
In other words, for large black holes $r_{\text{h}} \ll 1$ there is effectively no range for which the Araki--Lieb inequality is saturated, whereas for small black holes $r_{\text{h}} \gg 1$ the Araki--Lieb inequality is effectively always saturated. However, we may also consider intermediate regimes at which the transition happens at finite interval size, as shown in Figure \ref{figs:entPlatHP}.

Lastly, we emphasize that the Kasner exponent for any black hole solution to the vacuum equations of motion is given by a temperature-independent constant. For BTZ,
\begin{equation}
p_t^{\text{BTZ}} = 0.
\end{equation}
This can be seen by taking $d = 2$ in \eqref{kasnerVacuum}. The kinks thus constitute a line at $0$ in the parameter space $(\theta_{\text{k}},\phi_0,p_t)$ on the $\phi_0 = 0$ slice (at zero deformation parameter). In other words, the entanglement plateau transition does not imprint upon the interior geometry of the BTZ in the absence of matter.

\paragraph{Finite deformations} 
\begin{figure}
\centering
\includegraphics[scale=0.7]{figs/legendDeformations.pdf}
\vspace{-0.5cm}\\
\subfloat[$\theta_{\text{k}}$ versus $\beta$]{
\includegraphics[scale=0.55]{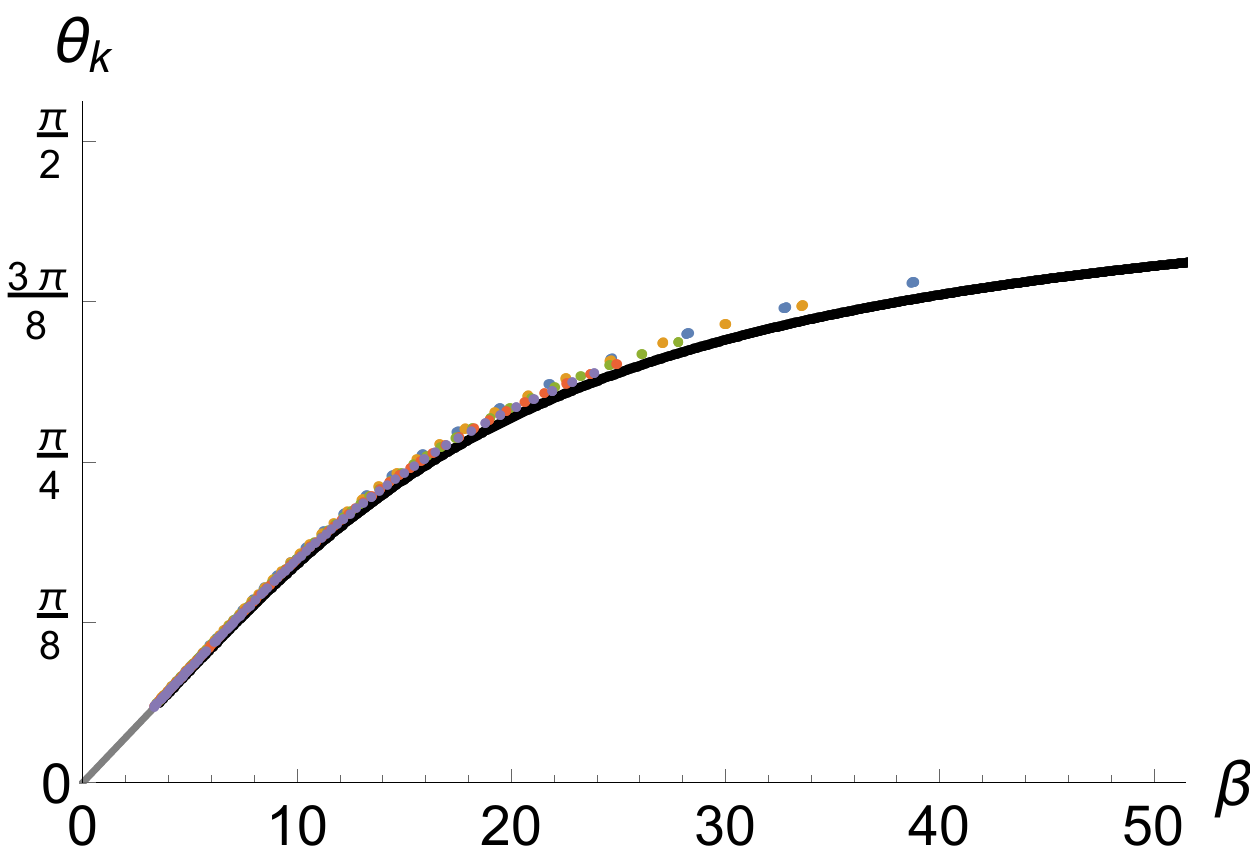}\label{fig5a}
}\ \ \ \ 
\subfloat[$p_t$ versus $\theta_{\text{k}}$]{
\includegraphics[scale=0.55]{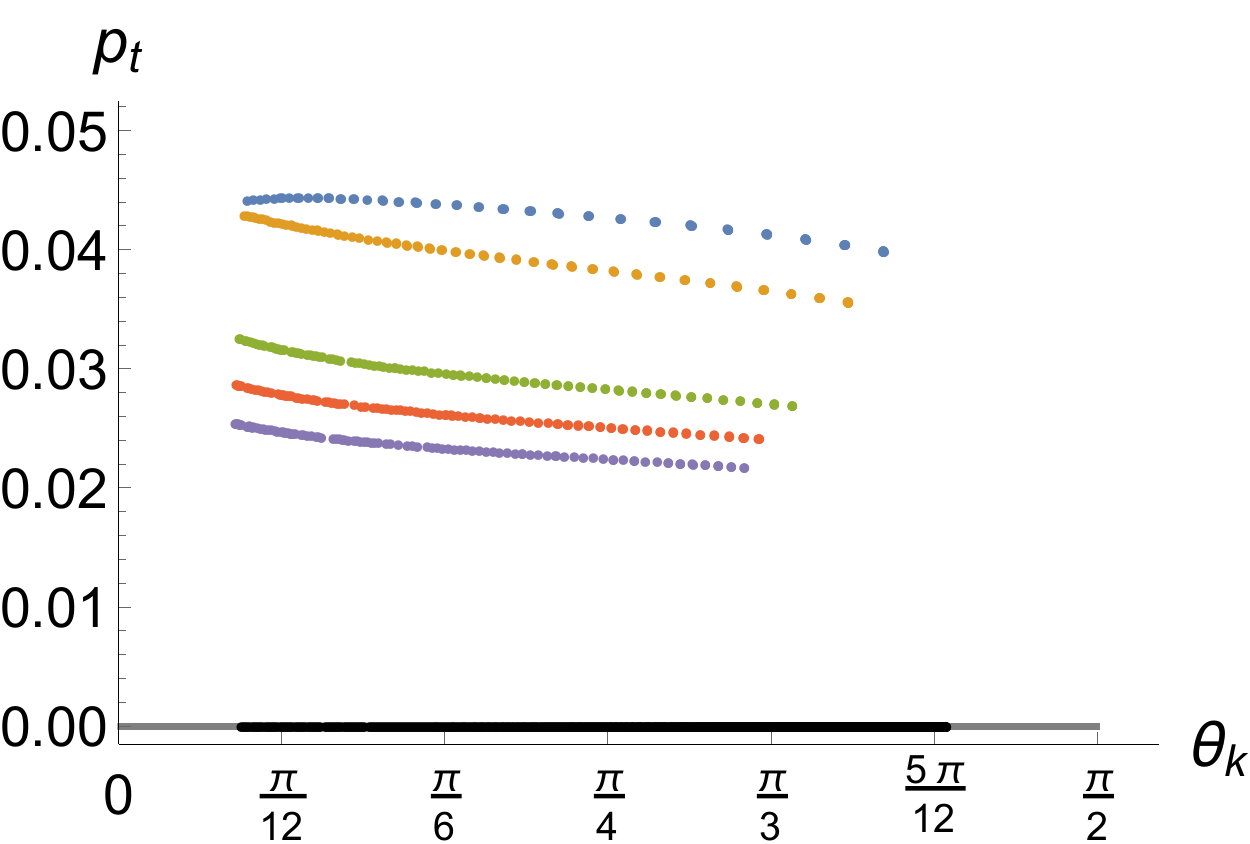}
\label{fig5b}}
\caption{The plots of (a) $\theta_{\text{k}}$ versus $\beta$ and (b) $p_t$ versus $\theta_{\text{k}}$ for a variety of scalar deformations characterized by the values of $\phi_0$. As our numerics do not find solutions for fixed $\phi_0$ outright, we compute these points while allowing for a $0.5\%$ margin of error in the values of $\phi_0$. As a sanity check, we numerically compute parameters from solutions with $\phi_0 \leq 0.005$ (close to $0$) and find that they indeed replicate the analytic results of the undeformed BTZ. The relationship between $\theta_{\text{k}}$ and $\beta$ appears insensitive to the deformation and shows strong matching to the analytic result for $\phi = 0$ \eqref{analyticKBTZBeta}. However, $p_t$ develops a nontrivial relationship with $\theta_{\text{k}}$ when $\phi_0 \not\approx 0$, in contrast to $p_t$ being uniformly $0$ when $\phi_0 \approx 0$.}
\label{figs:plotsPlateaux}
\end{figure}

For the BTZ, there is no relationship between $\theta_{\text{k}}$ (viewed as UV data) and the constant $p_t$ (the trans-IR data). Our goal now is to see whether this changes for finite values of the deformation parameter (in $\ell = 1$ units). This requires numerics.\footnote{Note that the conformal dimension $\Delta$ is another piece of UV data, but for our purposes we fix $\Delta = \frac{3}{2}$.} To summarize our results, we plot both $\theta_{\text{k}}$ versus $\beta$ and $p_t$ versus $\theta_{\text{k}}$ from points obtained via our numerical approach for a variety of deformations in Figure \ref{figs:plotsPlateaux}.

As validation, we first plot the points for $\phi_0 \leq 0.005$. These points should and do generally follow the analytic relations above, namely \eqref{analyticKBTZBeta} for $\theta_{\text{k}}$ versus $\beta$ and a flat line at $0$ for $p_t$ versus $\theta_{\text{k}}$. Subsequently, we then plot points for finite values of $\phi_0$, namely $\phi_0 = 5.000 \pm 0.025$, $10.00 \pm 0.05$, $50.00 \pm 0.25$, $100.0 \pm 0.5$, and $200. \pm 1$---allowing for errors of $0.5\%$ in our scan over deformation parameters.

Just as for the undeformed BTZ, $\theta_{\text{k}}$ monotonically increases in $\beta$, so we can swap $\theta_{\text{k}}$ for the temperature in the UV data. Interestingly, while we certainly observe a dependence of this curve on $\phi_0$, it appears to be very small. However, the relationship between $p_t$ and $\theta_{\text{k}}$ is much more subtle. Qualitatively, the plot appears stable against the $\pm 0.5\%$ error in $\phi_0$. Thus the numerics suggest that for finite, fixed deformations, $\theta_{\text{k}}$ has a numerically weak but nontrivial relationship with $p_t$. This is in contrast to no deformation being turned on ($\phi_0 \approx 0$), in which case $p_t$ is completely independent of $\theta_{\text{k}}$. The $\phi_0 \approx 5$ plot in particular suggests that the Kasner exponent peaks at some value of $\theta_{\text{k}}$. The higher values of $\phi_0$ are consistent with this behavior, assuming that their peaks appear at smaller values of $\theta_{\text{k}}$. 

These plots reflect how the UV entanglement plateau transition imprints upon the trans-IR data. Physically, both large and small values of $\theta_{\text{k}}$ appear to induce smaller values of $p_t$ within the range of our numerics. Thus, there is a particular black hole with maximal $p_t$ whose plateau occurs at an intermediate value of $\theta_{\text{k}}$. As per the interpretation of positive $p_t$ discussed in Section \ref{sec:dataTIR}, the interior geometry of this dual black hole has a maximally fast collapse relative to other states.

\section{Thermal two-point functions and the interior} \label{sec:thermal2pt}

Another entry of the holographic dictionary involving classical geometry is \textit{geodesic approximation} of correlation functions. In this section, we restrict our attention to scalar correlation functions, so we assume the presence of a second bulk scalar field $\tilde{\Phi}$ dual to some CFT operator $\tilde{\mathcal{O}}$ in the boundary theory with conformal dimension $\tilde{\Delta}$. Furthermore, for simplicity we assume that \textit{this} scalar field is not coupled to anything, so it does not generate any gravitational backreaction. We will eventually take $\tilde{\Delta}$ to be large (meaning that $\tilde{\mathcal{O}}$ will be irrelevant), so the presence of $\tilde{\Phi}$ stands in contrast to the scalar $\Phi$ in \eqref{actscalar} whose backreaction effects correspond to RG flow from the UV to the trans-IR.

The basic idea of geodesic approximation comes from seminal work equating the scalar propagator of a holographic CFT to a path integral over worldlines in the bulk \cite{Banks:1998dd,Balasubramanian:1999zv}. Specifically, taking two boundary points $\hat{x}_1$ and $\hat{x}_2$, the Euclidean scalar propagator is
\begin{equation}
G(\hat{x}_1,\hat{x}_2) = \int_{\hat{x}_1 \leftrightarrow \hat{x}_2} \mathcal{D}\mathcal{P}\,e^{-\tilde{\Delta}L[\mathcal{P}]}.\label{scalarpropbdry}
\end{equation}
Here, $\mathcal{P}$ is a generic bulk path connecting the boundary points, and $L[\mathcal{P}]$ is a length functional. Given a particular bulk geometry, the left-hand side is equivalent to the Euclidean two-point function of $\tilde{\mathcal{O}}$ in the corresponding CFT state.

Now, we take the limit $\tilde{\Delta} \to \infty$ by taking the bulk scalar mass to be large in units of AdS radius. Hereafter, we denote this large conformal dimension as $\Delta_{\text{H}}$ and the corresponding operator as $\mathcal{O}_{\text{H}}$. The path integral in \eqref{scalarpropbdry} may then be evaluated using saddle-point approximation. This picks out minima of the length functional---in other words, geodesics:
\begin{equation}
\expval{\mathcal{O}_{\text{H}}(\hat{x}_1)\mathcal{O}_{\text{H}}(\hat{x}_2)} \sim \sum_{\text{geodesics}} e^{-\Delta_{\text{H}} L}.
\end{equation}
We can actually go a bit further when working strictly in the $\Delta_{\text{H}} \to \infty$ limit. The sum above is dominated by the minimum-length saddle, and so among the different geodesics there is one in particular that actually determines the two-point function:
\begin{equation}
\expval{\mathcal{O}_{\text{H}}(\hat{x}_1)\mathcal{O}_{\text{H}}(\hat{x}_2)} \sim \exp\left[-\Delta_{\text{H}} \left(\stackrel[\text{geodesics}]{}{\text{min}} L\right)\right].
\end{equation}
With that said, we examine the geodesics in one-sided round black holes. This is equivalent to computing heavy \textit{thermal} two-point functions. We emphasize that the study of geodesics to glean insight into thermal correlation functions in CFT is not new \cite{Fidkowski:2003nf} and has been recently revitalized \cite{Grinberg:2020fdj,Berenstein:2022nlj}.

Our goal is to examine the phase structure of these heavy thermal two-point functions on compact spatial slices. In particular, we consider a CFT thermal state on $S^{d-1}$ and note that two operator insertions will always be collinear. Thus, the picture looks similar to Figure \ref{figs:roundEntanglement}, with two different connected geodesics being among the possible dominant phases of the two-point function. If either dominates the geodesic approximation, then we have a connected two-point function. In addition, there is a third candidate---a disconnected pair of geodesics, each of which is anchored to one of the insertions and hits the horizon. This third candidate represents the disconnected part of the two-point function (the product of the heavy thermal one-point functions \cite{Grinberg:2020fdj}). The three geodesics which contribute to the thermal two-point function are shown in Figure \ref{figs:round2pt}.

\input{figs/round2pt}

Just as in our exploration of the entanglement plateaux (Section \ref{sec:plateaux}), the discussion in this paper will be focused on $d = 2$. We do so for simplicity, but note that unlike codimension-2 entanglement surfaces, geodesics are always strictly 1d regardless of the number of spatial dimensions. The main differences in higher dimensions are the metric functions and the space of gravitational solutions.\footnote{In higher dimensions, we have both small and large black holes of different horizon radii but at equal temperatures. Furthermore, there is a strict upper bound on $\beta$ for the regime in which there exist black hole solutions.} Therefore, we focus on $d = 2$ to simplify the picture, leaving these additional complications to future work.

One of the useful simplifications of working in $d = 2$ is that geodesics are also codimension-2, so we can and will employ many of the same equations as in Section \ref{sec:plateaux}. We will also present the story in a similar manner. This time, however, we will find a connected/disconnected phase transition of the heavy thermal two-point function and show how its characteristic parameter imprints upon the near-singularity geometry.

We also note that the story here has a similar flavor to the story of holographic confinement/deconfinement phase transitions (cf. \cite{Maldacena:1998im,Klebanov:2007ws}). However, while we are employing similar geometrical structures (lines), the phase transition we examine is different.

\subsection{Geodesic transition in deformations of the BTZ black hole}

We reiterate that the general $d = 2$ metric (setting $\ell = 1$) describing deformations of the round BTZ black hole is
\begin{equation}
ds^2 = \frac{1}{r^2}\left[-e^{-\chi(r)} F(r)dt^2 + \frac{dr^2}{F(r)} + d\theta^2\right],
\end{equation}
Without loss of generality, we consider boundary insertions on the same time slice---one at $\theta = \theta_{\mathcal{O}}$ and the other at $\theta = -\theta_{\mathcal{O}}$, where $\theta_{\mathcal{O}} \in (0,\pi)$ denotes half of the angular separation between the insertions. This is precisely how we defined the entanglement subregion in Section \ref{sec:plateaux}. The two-point function is
\begin{equation}
\expval{\mathcal{O}_{\text{H}}(\hat{x}_1)\mathcal{O}_{\text{H}}(\hat{x}_2)} = \expval{\mathcal{O}_{\text{H}}(\theta_{\mathcal{O}})\mathcal{O}_{\text{H}}(-\theta_{\mathcal{O}})}.
\end{equation}

\paragraph{Lengths of different phases} Now recall that the entanglement surfaces of $d = 2$ are precisely geodesics. Thus, the generic expressions for the geodesic lengths are obtained from rebranding \eqref{eq:rc}--\eqref{areaCbh} and \eqref{eq:rd}--\eqref{areaDbh}:
\begin{align}
L_{\text{C1}} &= 2\int_0^{r_{\text{c1}*}} \frac{dr}{\sqrt{F(r)}} \frac{r_{\text{c1}*}/r}{\sqrt{r_{\text{c1}*}^2-r^2}}, \ \ \theta_{\mathcal{O}} = \int_0^{r_{\text{c1}*}} \frac{dr}{\sqrt{F(r)}}\frac{r}{\sqrt{r_{\text{c1}*}^2 - r^2}},\label{barec1}\\
L_{\text{C2}} &= 2\int_0^{r_{\text{c2}*}} \frac{dr}{\sqrt{F(r)}} \frac{r_{\text{c2}*}/r}{\sqrt{r_{\text{c2}*}^2-r^2}}, \ \ \theta_{\mathcal{O}} = \pi - \int_0^{r_{\text{c2}*}} \frac{dr}{\sqrt{F(r)}}\frac{r}{\sqrt{r_{\text{c2}*}^2 - r^2}}.\label{barec2}
\end{align}
Here we use C1 to represent the connected geodesic which does not go around the horizon. Meanwhile, C2 denotes the connected geodesic which wraps the horizon.

We must also consider the disconnected term---that is, the product of thermal one-point functions. In Euclidean signature, this is computed by a pair of geodesics for which $\theta'(r) = 0$ and that both reach the horizon \cite{Grinberg:2020fdj}. The total length of these geodesics is
\begin{equation}
L_{\text{D}} = 2\int_{0}^{r_{\text{h}}} \frac{dr}{r\sqrt{F(r)}}.\label{bared}
\end{equation}

\paragraph{Renormalized lengths} The lengths above diverge at the conformal boundary. To renormalize them, we use $-L_{\text{D}}$ as a local counterterm, which of course means that the renormalized disconnected geodesic length $L_{\text{D}}^{\text{ren}}$ is set to $0$. Noting that $r_{\text{h}} > r_{\text{c1}*}, r_{\text{c2}*}$, we can write the renormalized connected geodesic lengths as sums of manifestly finite integrals:
\begin{align}
L_{\text{C1}}^{\text{ren}} &= 2\int_{0}^{r_{\text{c1}*}} \frac{dr}{r\sqrt{F(r)}} \frac{r_{\text{c1}*} - \sqrt{r_{\text{c1}*}^2 - r^2}}{\sqrt{r_{\text{c1}*}^2 - r^2}} - 2\int_{r_{\text{c1}*}}^{r_{\text{h}}} \frac{dr}{r\sqrt{F(r)}},\\
L_{\text{C2}}^{\text{ren}} &= 2\int_{0}^{r_{\text{c2}*}} \frac{dr}{r\sqrt{F(r)}} \frac{r_{\text{c2}*} - \sqrt{r_{\text{c2}*}^2 - r^2}}{\sqrt{r_{\text{c2}*}^2 - r^2}} - 2\int_{r_{\text{c2}*}}^{r_{\text{h}}} \frac{dr}{r\sqrt{F(r)}}.
\end{align}
This approach is rather friendly for our numerics, since the explicit form of the blackening function $F(r)$ is unknown and the integrals cannot be evaluated analytically. However, when simply considering the BTZ, we may alternatively introduce a cutoff at $r = \epsilon$ and regulate the bare integrals \eqref{barec1}--\eqref{bared}.

\subsection{Transition angles versus Kasner exponents}

We now have the necessary equations to study the phase structure of the heavy thermal two-point correlator. We will discuss the numerical results shortly, but first we address the case of the undeformed BTZ black hole so as to obtain some analytic results.

\paragraph{Undeformed BTZ}

Recall the blackening function of the BTZ with no backreaction is
\begin{equation}
F_{\text{BTZ}}(r) = 1 - \frac{r^2}{r_{\text{h}}^2}.
\end{equation}
We already have the equations for the \textit{regulated} (with some cutoff at $r = \epsilon$) connected geodesic lengths; they are read from \eqref{connected} and \eqref{disconnected} as
\begin{align}
L_{\text{C1},\text{BTZ}}^{\text{reg}} &= -2 \log\left(\frac{\epsilon}{2r_h}\right) + \log\left(\frac{r_{\text{c1}*}^2}{r_{\text{h}}^2 - r_{\text{c1}*}^2}\right),\\
L_{\text{C2},\text{BTZ}}^{\text{reg}} &= -2\log\left(\frac{\epsilon}{2r_h}\right) + \log\left(\frac{r_{\text{c2}*}^2}{r_{\text{h}}^2 - r_{\text{c2}*}^2}\right),
\end{align}
where $r_{\text{c1}*}$ is identified with \eqref{connectedr} while $r_{\text{c2}*}$ is identified with \eqref{disconnectedr} (swapping $\theta_{\mathcal{R}}$ for $\theta_{\mathcal{O}}$). Furthermore, the regulated disconnected geodesic length is straightforward to calculate. It is simply
\begin{equation}
L_{\text{D},\text{BTZ}}^{\text{reg}} = \log\left(\frac{r_{\text{h}} + \sqrt{r_{\text{h}}^2 - \epsilon^2}}{r_{\text{h}} - \sqrt{r_{\text{h}}^2 - \epsilon^2}}\right) = -2\log\left(\frac{\epsilon}{2r_{\text{h}}}\right) + O(\epsilon).
\end{equation}
Hence, the renormalized lengths in terms of $\theta_{\mathcal{O}}$ are
\begin{align}
L_{\text{C1},\text{BTZ}}^{\text{ren}} &= 2\log\sinh\left(\frac{\theta_{\mathcal{O}}}{r_{\text{h}}}\right),\\L_{\text{C2},\text{BTZ}}^{\text{ren}} &= 2\log\sinh\left(\frac{\pi-\theta_{\mathcal{O}}}{r_{\text{h}}}\right),\\
L_{\text{D},\text{BTZ}}^{\text{ren}} &= 0.
\end{align}
First, note that the connected geodesics match when $\theta_{\mathcal{O}} = \frac{\pi}{2}$ and are symmetric around this point. This is true regardless of the temperature. Furthermore, they exchange dominance at this point:
\begin{figure}
\centering
\includegraphics[scale=0.75]{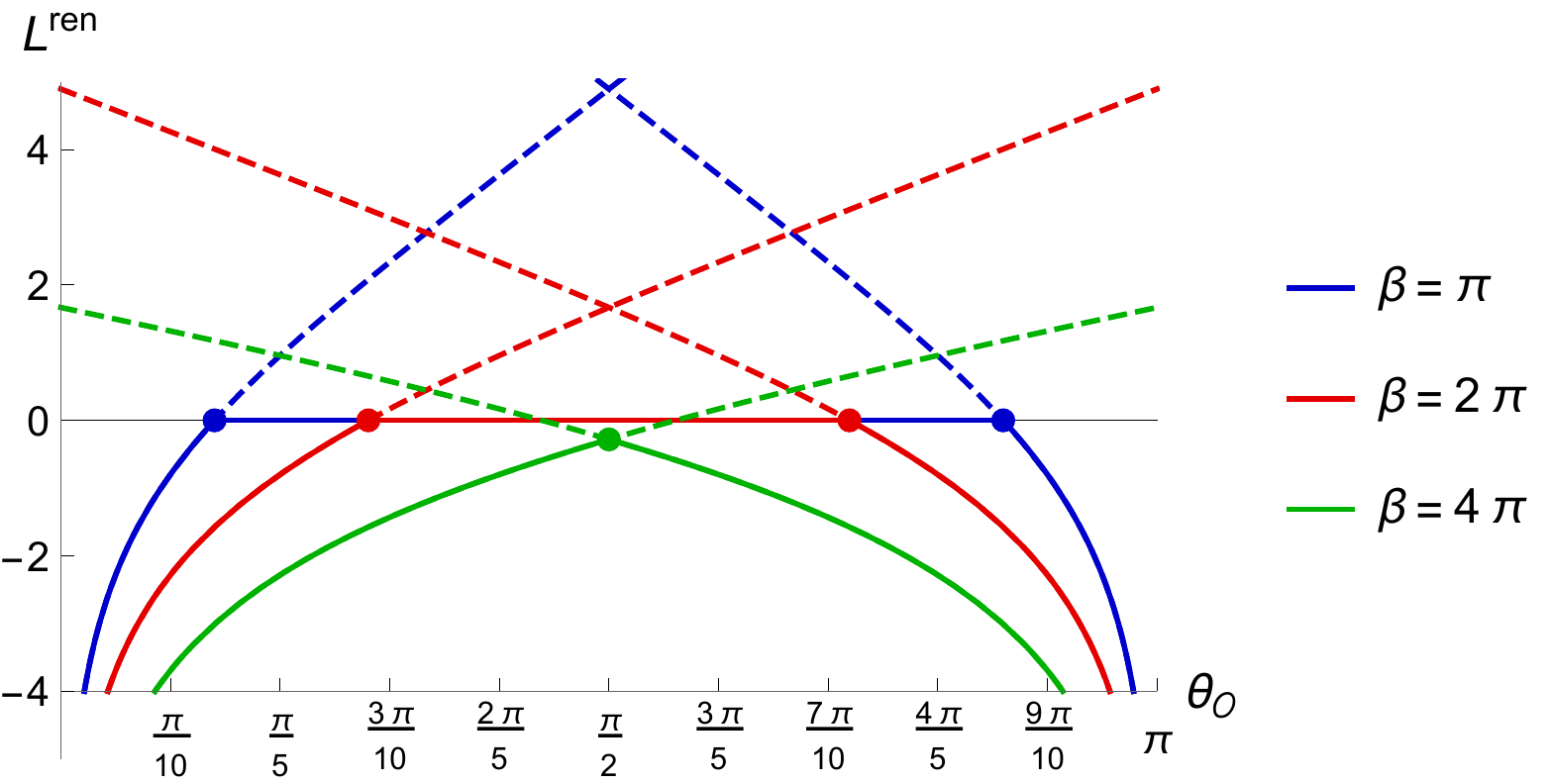}
\caption{The geodesic lengths for the BTZ black hole (with $\ell = 1$) for inverse temperatures $\beta = \pi$ (blue), $\beta = 2\pi$ (red), and $\beta = 4\pi$ (green). For each temperature, the solid lines follow the minimal length dominating the geodesic approximation of the heavy thermal two-point function. Dashed curves are the lengths of subleading connected geodesics. For the blue and red curves, \eqref{betarange2pt} is satisfied, so there is a transition between connected and disconnected phases. However, the green curve does not exhibit such a transition because there is always a connected geodesic with negative renormalized length that dominates the disconnected geodesic.}
\label{figs:BTZGeodesics}
\end{figure}
\begin{align}
\theta_{\mathcal{O}} < \frac{\pi}{2} &\implies L_{\text{C1},\text{BTZ}} < L_{\text{C2},\text{BTZ}},\\
\theta_{\mathcal{O}} > \frac{\pi}{2} &\implies L_{\text{C1},\text{BTZ}} > L_{\text{C2},\text{BTZ}}.
\end{align}
This is a first-order (continuous and not smooth) transition, but it may or may not be relevant to leading order in geodesic approximation. To make this statement more precise, we focus on the $\theta_{\mathcal{O}} < \frac{\pi}{2}$ regime. We observe that the dominant connected geodesic (C1) negatively diverges as $\theta_{\mathcal{O}} \to 0^+$ but reaches a finite value at $\frac{\pi}{2}$. This value may be positive or negative depending on the horizon radius $r_{\text{h}}$. Specifically,
\begin{align}
L_{\text{C1},\text{BTZ}}^{\text{ren}}\left(\theta_{\mathcal{O}} = \frac{\pi}{2}\right) = 2\log\sinh\left(\frac{\pi}{2r_{\text{h}}}\right) \begin{cases}
> 0&\text{if}\ \ r_{\text{h}} < \dfrac{\pi}{2\log(1 + \sqrt{2})},\vspace{0.2cm}\\
< 0&\text{if}\ \ r_{\text{h}} > \dfrac{\pi}{2\log(1 + \sqrt{2})}.
\end{cases}
\end{align}
This means that if the black hole is big enough (small $r_{\text{h}}$), then the connected geodesic C1 becomes larger than the disconnected geodesic D \textit{before} $\theta_{\mathcal{O}} = \frac{\pi}{2}$. We thus get a connected/disconnected phase transition. However, if the black hole is small, then this transition does not happen at leading order in the geodesic approximation.

Let us write the transition as a function of the inverse temperature $\beta = 2\pi r_{\text{h}}$. The connected/disconnected transition occurs at the point where
\begin{equation}
L_{\text{C1},\text{BTZ}}^{\text{ren}} = 0 \implies \theta_{\mathcal{O}} = \theta_{\text{CD}}^{\text{BTZ}} \equiv r_{\text{h}}\log(1 + \sqrt{2}) = \frac{\beta}{2\pi}\log(1 + \sqrt{2}),\label{analyticCDang}
\end{equation}
where $\theta_{\text{CD}}^{\text{BTZ}}$ is used to denote the transition angle for a particular BTZ black hole. However, the occurrence of this transition is restricted to the range
\begin{equation}
\theta_{\text{CD}}^{\text{BTZ}} < \frac{\pi}{2} \iff \beta < \frac{\pi^2}{\log(1 + \sqrt{2})} \approx 3.56\pi.\label{betarange2pt}
\end{equation}
With the above relation in mind, we plot the renormalized geodesic lengths for three values of $\beta$ in Figure \ref{figs:BTZGeodesics} so as to depict the phase structure of the heavy thermal two-point function.

Just as for the entanglement plateau transition, this connected/disconnected transition does not imprint upon the near-singularity geometry of the BTZ black hole; the Kasner exponent is fixed at $0$ regardless of the value of $\theta_{\text{CD}}^{\text{BTZ}}$. We now explore how this changes when backreaction effects are turned on.

\paragraph{Finite deformations}

\begin{figure}
\centering
\includegraphics[scale=0.7]{figs/legendDeformations.pdf}
\vspace{-0.5cm}\\
\subfloat[$\theta_{\text{CD}}$ versus $\beta$]{
\includegraphics[scale=0.55]{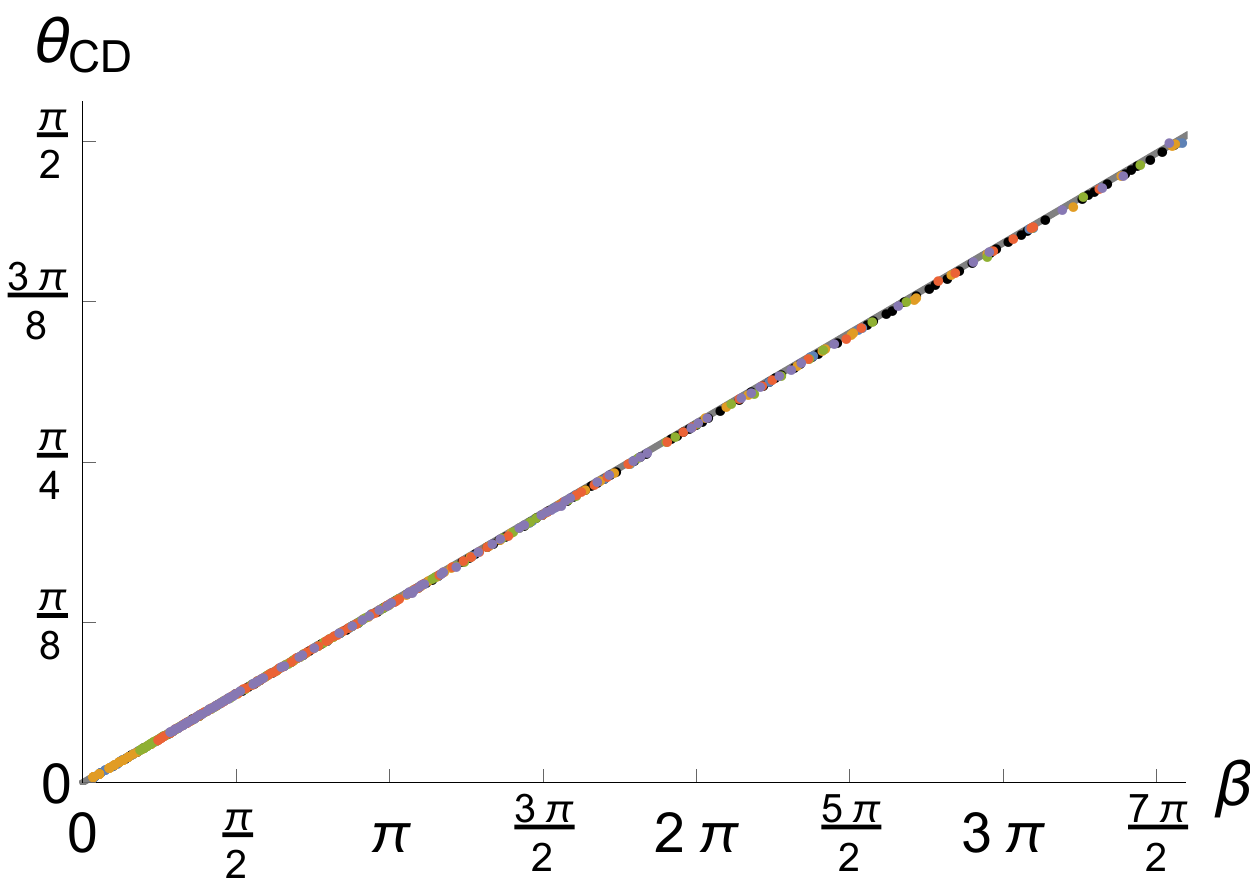}\label{fig8a}
}\ \ \ \ 
\subfloat[$p_t$ versus $\theta_{\text{CD}}$]{
\includegraphics[scale=0.55]{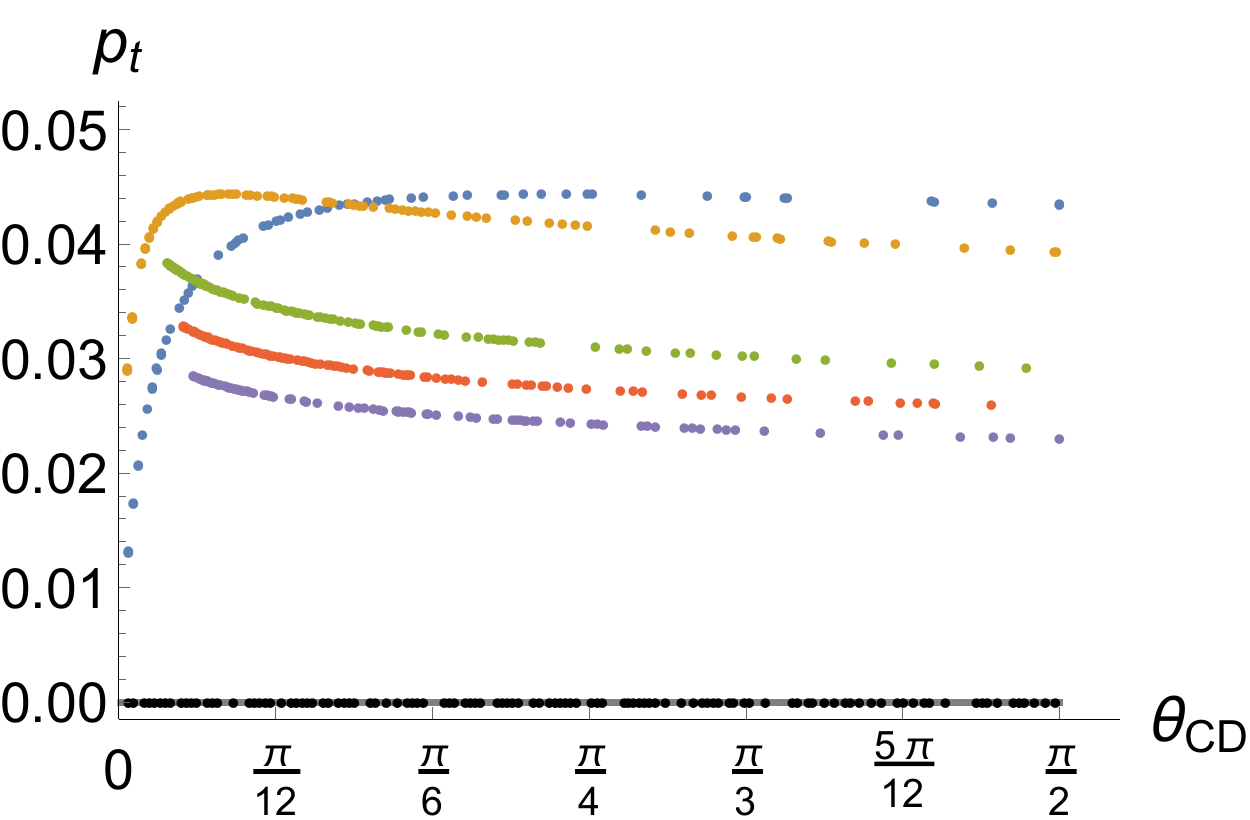}
}
\caption{The plots of (a) $\theta_{\text{CD}}$ versus $\beta$ and (b) $p_t$ versus $\theta_{\text{CD}}$ for a variety of scalar deformations characterized by the values of $\phi_0$, with $\theta_{\text{CD}}$ restricted to $\left(0,\frac{\pi}{2}\right)$. We again compute these points while allowing for a $0.5\%$ margin of error in the values of $\phi_0$ (or a maximal error of $0.005$ for $\phi_0 \approx 0$). $(\text{a})$ We see the analytic result of the underformed BTZ reproduced by the $\phi_0 \approx 0$ points. We also observe that the relationship between $\theta_{\text{CD}}$ and $\beta$ appears insensitive to the deformation. $(\text{b})$ Thus, $p_t$ develops a nontrivial relationship with $\theta_{\text{CD}}$ when $\phi_0 \not\approx 0$ which is the same shape as the $p_t$ versus $\beta$ plot of Figure \ref{figs/ptvsbeta-2d}.}
\label{figs:finiteDefCD}
\end{figure}

We again look at what happens when $\phi_0 \not\approx 0$ using numerics. The procedure is the same as for the entanglement plateaux; we first construct a large set of solutions and then filter them out to take only those which are within $0.5\%$ of the target value of $\phi_0$. As a calibration check, we also confirm that the numerics of the solutions for which $\phi_0 \approx 0$ reproduce the analytic relationship \eqref{analyticCDang} between $\theta_{\text{CD}}$ and $\beta$, along with the upper bound \eqref{betarange2pt}.

The resulting plots are shown in Figure \ref{figs:finiteDefCD}. Interestingly, our numerics suggest that the linear relationship between $\theta_{\text{CD}}$ and $\beta$ found in the undeformed case is completely insensitive to $\phi_0$. The deformed geometries appear to give the same exact line, implying that $\theta_{\text{CD}}$ is a well-defined UV datum. As such, the relationship between $p_t$ and $\theta_{\text{CD}}$ is qualitatively similar to the one shown in Figure \ref{figs/ptvsbeta-2d} between $p_t$ and $\beta$; $p_t$ exhibits an apparent maximum in $\beta$ when $\phi_0 \not\approx 0$.

This imprint is numerically much stronger than what we had seen from the entanglement plateau transition. Mathematically, this is because here we have a linear relationship between $\theta_{\text{CD}}$ and $\beta$, whereas the critical size $\theta_{\text{k}}$ exhibits something akin to a logistic growth with $\beta$ (Figure \ref{fig5a}). Nonetheless, the qualitative story is similar; for each $\phi_0$, there is a particular intermediate value of $\theta_{\text{CD}}$ at which the interior geometry of the black hole has a maximally fast collapse due to the Kasner exponent being peaked. The difference is that the peak is much more obvious in the relationship between $p_t$ and $\theta_{\text{CD}}$ than in Figure \ref{fig5b}.

\section{Discussion} \label{sec:discussion}

This paper is guided by the claim that the interior of a black hole can be viewed as an analytically continued RG flow. This is a working assumption that comes out of attempting to synthesize the holographic RG flow perspective on the emergence of spacetime with the existence of a change in the radial extra dimension's signature from spacelike to timelike at the horizon. In previous work, two of the authors had supplemented this claim in the cases of flat isotropic \cite{Caceres:2022smh} and anisotropic \cite{Caceres:2022hei} black holes by constructing a monotone from the null energy condition. In Section \ref{sec:roundBH}, we have performed the analogous construction in the round black hole as a motivation for its analysis.

We have used the round black hole deformed by a scalar as a model system for how imprints of UV phase transitions persist near the classical Kasner singularity, which is viewed as the ``endpoint" of the trans-IR flow. Much of the study was motivated by similar work \cite{Hartnoll:2020fhc} in the holographic superconductor. In the present work, we have focused on phase transitions holographically dual to transitions in bulk geometrical structures---namely those of entanglement surfaces (Section \ref{sec:plateaux}) and geodesics (Section \ref{sec:thermal2pt}). Such transitions are universal in that they do not depend on the specific matter content of the bulk theory. Our results show that the parameters characterizing both phase transitions develop a nontrivial relationship with the Kasner exponent---characterizing the singularity---only when a deformation is turned on. This contrasts with the story in black holes solutions without matter, for which the Kasner singularity is only dependent on the number of spacetime dimensions and no UV physics.

\subsection*{Future directions}

There are several ways to proceed from here. We list some of these follow-up ideas below.

\paragraph{The Hawking--Page transition}

In this paper, we consider a CFT at finite temperature with states on a spatial $(d-1)$-sphere organized in a canonical ensemble. The dual states in the bulk are either black holes or thermal gases of gravitons in AdS$_{d+1}$. The ensemble of gravitational states is known to exhibit a Hawking--Page transition \cite{Hawking:1982dh} at some fixed temperature, which with no matter present is \cite{Witten:1998zw}
\begin{equation}
\beta_{\text{HP}} = \frac{2\pi}{d-1}.
\end{equation}
Specifically, the partition function is dominated by the state dual to the bulk saddle with the smallest on-shell action. The leading-order thermal state at temperature $\beta$ is typically the large black hole when $\beta < \beta_{\text{HP}}$ and the thermal gas when $\beta > \beta_{\text{HP}}$.

A very early entry in the AdS/CFT dictionary \cite{Witten:1998zw} is the statement that this Hawking--Page transition corresponds to a confinement/deconfinement phase transition on the level of the free energy of the dual CFT or gauge theory. One can also add matter to the bulk picture (cf. \cite{Gursoy:2010jh,Gursoy:2018umf}), and furthermore the transition seems to not exist for flat or hyperbolic horizon topologies \cite{Birmingham:1998nr,Cho:2002hq}. In other words, the Hawking--Page transition is also characteristic to the round black hole, just like the transitions we consider.

It would be interesting to understand how the Hawking--Page transition imprints upon the near-singularity Kasner geometry. However, this would require understanding the phase structure of the canonical ensemble. While the relative stability of the hairy versus hairless phases have been studied (cf. \cite{Hertog:2004bb}), we propose fixing the source $\phi_0$ rather than treating it as a variable. The resulting phase diagram would include a Hawking--Page transition scale $\beta_{\text{HP}}$ and, when $d > 2$, a spinodal scale $\beta_{\text{sp}}$ above which no black holes exist. Once these scales are known for a particular $\phi_0$, one may then construct plots of $p_t$ versus $\beta$ as in Figure \ref{figs:ptbetaplots} to say what happens to the near-singularity Kasner geometry as the temperature approaches either scale.\footnote{We can already say what happens at the spinodal scale in the $d > 2$ case---the small and large black holes' values for $p_t$ approach one another. We emphasize that this is completely expected because the spinodal point is where the two types of black holes become geometrically identical.}

\paragraph{Higher dimensions} In studying the imprints of the entanglement plateau transition on the trans-IR data, we focused on black holes in $d = 2$ because they are much more tractable in comparison to the higher-dimensional ($d > 2$) case. We did the same for the connected/disconnected transition of the thermal two-point function. However, the $d = 2$ scalar-deformed black holes typically have positive Kasner exponents, indicating collapsing Einstein--Rosen bridges, whereas $d > 2$ scalar-deformed black holes have \textit{negative} Kasner exponents. In other words, the $d > 2$ case is both more generic and involves more ``stable" Einstein--Rosen bridges when compared to $d = 2$.

It would be satisfying to test our core assertion that UV physics imprints nontrivially upon trans-IR geometry in the more generic case. It is somewhat reasonable to expect that the story of the thermal two-point function does not change much, since the geometric objects being considered are always geodesics. However, it is possible that our assertion may fail for the entanglement plateaux transition in $d > 2$. As entanglement in AdS/CFT is not a probe of the singularity due to the existence of extremal surface barriers \cite{Engelhardt:2013tra}, such an expectation would be valid.

Nonetheless, actually checking the higher-dimensional case would require a more sophisticated numerical approach than what has been done here, particularly if we were to study the entanglement plateaux for scalar-deformed black holes. One idea is to construct the higher-dimensional black holes numerically and then apply mean curvature flow to compute entanglement surfaces for boundary caps, as was done in \cite{Hubeny:2013gta} for AdS black holes without matter.

\paragraph{UV physics and singularity dynamics} As mentioned in the main text (Section \ref{sec:dataTIR}), the story of the classical near-singularity geometry of black holes in general relativity is old. There have been many models in the literature put forth as ways to think about features of such geometry. One of the seminal models goes under the name of ``cosmological billiards" \cite{Damour:2002et,Damour:2002tc} and provides a simple perspective on dynamics near the singularity. There has also been another model \cite{Hartnoll:2022rdv} utilizing a superexponential scalar potential to exercise control over Kasner ``oscillations" in which $p_t$ flips between different values all the way to the singularity. It would be interesting to connect UV physics to these pictures of the trans-IR regime.

\acknowledgments 

We thank Andreas Karch, Juan Pedraza and Mukund Rangamani for useful discussions. EC and SS are supported by the National
Science Foundation under Grant No. PHY-2112725. SS is also supported by NSF Grant No. PHY-1914679. HS is supported by a grant from the Simons Foundation (Grant 651440, AK). EC thanks the Instituto de F\'isica Te\'orica at Universidad Aut\'onoma de Madrid, for hospitality during the last stages of this project. SS likewise acknowledges the Santa Cruz Institute of Particle Physics at the University of California, Santa Cruz for hospitality towards the end of this project. Finally, HS thanks the Ultra-Quantum Matter collaboration meeting at the University of Colorado Boulder for hospitality when the draft was being finalized.

\begin{appendices}

\section{Holographic $a$-theorem for global AdS deformations}\label{app:pureAdS} 

Asymptotically AdS$_{d+1}$ spacetimes with planar boundary and without a horizon, i.e., those corresponding to zero-temperature states, can be written in the form
\begin{equation}
\label{eq:cylinder}
ds^2 = e^{2A(\rho)} \left(-dt^2 + d\vec{x}_{d-1}^2\right) + d\rho^2,
\end{equation}
where $(t,\vec{x}_{d-1}) \in \mathbb{R}^d$ and $\rho \in \mathbb{R}$, with $\rho = \infty$ being the conformal boundary and $\rho = -\infty$ being the Poincar\'e horizon. $A$ is a real function controlling the backreaction, and empty AdS with radius $\ell$ itself corresponds to $A(\rho) = \frac{\rho}{\ell}$. We take the geometry to be asymptotically AdS with radius $\ell$,
\begin{equation}
A(\rho) \sim \frac{\rho}{\ell},\ \ \rho \to \infty.
\end{equation}
Recalling \cite{Freedman:1999gp}, the holographic $a$-function of such spacetimes constructed from the holographic trace anomaly coefficient $\sim \ell^{d-1}$ \cite{Henningson:1998gx} and the null energy condition is\footnote{Note that we put the topology of the boundary spatial slice in the function's argument. In this appendix, we do so to distinguish the $a$-function for a flat slicing from that of a round slicing.}
\begin{equation}
a(\rho;\mathbb{R}^{d-1}) = \frac{\pi^{d/2}}{\Gamma\left(\frac{d}{2}\right)\ell_{\text{P}}^{d-1}} \left[\frac{1}{A'(\rho)}\right]^{d-1}.\label{flatafunction}
\end{equation}
Our goal here is to construct the analogous function for a cylindrical boundary topology. These constitute deformations of global AdS$_{d+1}$ and describe RG flows of states on the spatial sphere $S^{d-1}$. More concretely, consider the domain-wall ansatz
\begin{equation}
ds^2 = e^{2A(\rho)} \left[-dt^2 + \alpha^2 e^{2\mathcal{R}(\rho)} d\Omega_{d-1}^2\right] + d\rho^2,\label{eq:roundempty}
\end{equation}
where $t \in \mathbb{R}$, $\Omega_{d-1}$ is the line element of a $(d-1)$-sphere, and $\rho \geq 0$. While $\rho = \infty$ is still the conformal boundary, $\rho = 0$ is the central axis of the AdS cylinder.\footnote{There is no Poincar\'e horizon in the AdS cylinder.} $\alpha$ is a (for now) free length scale describing the radius of the boundary $S^{d-1}$, and $\mathcal{R}$ is a real function determining the radius of each fixed-$\rho$ slice. We get empty AdS$_{d+1}$ with curvature radius $\ell$ if
\begin{equation}
A(\rho) = \log\cosh\left(\frac{\rho}{\ell}\right),\ \ \mathcal{R}(\rho) = \log\tanh\left(\frac{\rho}{\ell}\right),\ \ \alpha = \ell.\label{functionsGlobalAdS}
\end{equation}
Thus, bulk deformations of empty AdS$_{d+1}$ with radius $\ell$ must still have functions $A,\mathcal{R}$ and length scale $\alpha$ which match these profiles asymptotically. This assertion fixes $\alpha$ in the \textit{entire} geometry to be $\ell$. Furthermore, at $\rho \to \infty$ the metric becomes
\begin{equation}
ds^2 \sim \frac{e^{2\rho/\ell}}{4} \left(-dt^2 + \ell^2 d\Omega_{d-1}^2\right),
\end{equation}
so the UV CFT lives on a cylinder imbued with the flat metric and of radius $\ell$.

With all of that in mind, we are now ready to demonstrate that the natural holographic $a$-function condition is
\begin{equation}
a(\rho;S^{d-1}) = \frac{\pi^{d/2}}{\Gamma\left(\frac{d}{2}\right)\ell_{\text{P}}^{d-1}} \left[\frac{e^{-\mathcal{R}(\rho)}}{A'(\rho) + \mathcal{R}'(\rho)}\right]^{d-1}.\label{sphereafunction1}
\end{equation}
We first focus on how the $\rho$ dependence in this expression is derived and shown to be monotonic from the null energy condition. We then confirm that \eqref{sphereafunction1} properly reduces to the holographic trace anomaly coefficient in the UV.

\subsection{Construction for round slices without horizon}

Our first step is to reverse engineer an $a$-function from the ``radial" null energy condition. In Einstein gravity, this takes the form
\begin{equation}
T_{\mu\nu}k^\mu k^\nu \geq 0,\quad
\ell_{\text{P}}^{d-1} T_{\mu\nu} = G_{\mu\nu} - \frac{d(d-1)}{\ell^2} g_{\mu\nu},\quad
k^\mu = e^{-A(\rho)} \delta^{\mu}_t + \delta^{\mu}_{\rho}.\label{necvac}
\end{equation}
We use the scheme outlined in \cite{Caceres:2022hei}---namely, to write the contraction of this stress tensor in the form
\begin{equation}
T_{\mu\nu}k^\mu k^\nu = \mathcal{C}(\rho)\frac{d}{d\rho}\left[\mathfrak{a}(\rho)^{1/(d-1)}\right] - \mathcal{K}(\rho)^2,\label{schemeT}
\end{equation}
for some positive function $\mathcal{C}$ and real functions $\mathfrak{a}$ and $\mathcal{K}$. If we can do so and also prove that
\begin{equation}
\mathfrak{a}(\rho)^{(d-2)/(d-1)} > 0,\label{ineqascheme}
\end{equation}
then the null energy condition \eqref{necvac} ensures that $a(\rho)$ monotonically increases as $\rho$ increases. Indeed, we find that the schematic relation \eqref{schemeT} is satisfied with
\begin{equation}
\mathcal{C}(\rho) = \frac{(d-1)}{\ell_{\text{P}}^{d-1}}e^{\mathcal{R}(\rho)} \left[A'(\rho) + \mathcal{R}'(\rho)\right]^2,\quad
\mathcal{K}(\rho) = 0,\quad
\mathfrak{a}(\rho) = \left[\frac{e^{-\mathcal{R}(\rho)}}{A'(\rho) + \mathcal{R}'(\rho)}\right]^{d-1}.
\end{equation}
Furthermore, \eqref{ineqascheme} is true if
\begin{equation}
A'(\rho) + \mathcal{R}'(\rho) > 0.\label{ineqsum}
\end{equation}
We can indeed show this to be the case by using the null energy condition and doing a proof by contradiction. Specifically, we may assume that there exists a $\rho = \rho_*$ at which $A' + \mathcal{R}'$ vanishes and above which $A' + \mathcal{R}'$ is positive. Such a point may generally exist because $A' + \mathcal{R}'$ asymptotes to a positive number $\ell^{-1}$ by \eqref{functionsGlobalAdS}. Thus, we may write this sum and its derivative as Taylor series of the form
\begin{align}
&A'(\rho) + \mathcal{R}'(\rho) = c_*(\rho - \rho_*)^{p_*} + O[(\rho-\rho_*)^{p_*+1}],\\
&A''(\rho) + \mathcal{R}''(\rho) = c_* p_* (\rho-\rho_*)^{p_*-1} + O[(\rho-\rho_*)^{p_*}],
\end{align}
where $c_* > 0$ and $p_* \geq 1$. Thus for some small $\epsilon > 0$, we may approximate the left-hand sides as
\begin{align}
&A'(\rho_* + \epsilon) + \mathcal{R}'(\rho_* + \epsilon) \approx c_* \epsilon^{p_*},\\
&A''(\rho_* + \epsilon) + \mathcal{R}''(\rho_* + \epsilon) \approx c_* p_* \epsilon^{p_*-1}.
\end{align}
Now we may approximate $A'(\rho_* + \epsilon)$ and $A''(\rho_* + \epsilon)$. This allows us to write the contraction of the stress tensor \eqref{schemeT} to leading order in small $\epsilon$:
\begin{equation}
\left.T_{\mu\nu}k^\mu k^\nu\right|_{\rho = \rho_* + \epsilon} \approx -(d-1)c_* p_* \epsilon^{p_* - 1}.\label{approxGlobal}
\end{equation}
The punchline is that this expression is manifestly negative as we approach $\rho = \rho_*$ from above, in contradiction with the null energy condition. This completes the argument, so we may say that \eqref{ineqsum} holds. This in turn means that
\begin{equation}
\mathfrak{a}(\rho) = \left[\frac{e^{-\mathcal{R}(\rho)}}{A'(\rho) + \mathcal{R}'(\rho)}\right]^{d-1}\label{monFunc}
\end{equation}
monotonically increases with $\rho$. For a black hole, the analogous candidate looks similar up to a blackening factor in the numerator [cf. \eqref{aCandBH}].

\subsection{Validation in the UV} 

There is one small test of our proposed $a$-function that we need to perform for validation; we must ensure that \eqref{sphereafunction1} reduces to the usual holographic trace anomaly coefficient at the conformal boundary. This just amounts to plugging \eqref{functionsGlobalAdS} into \eqref{monFunc} (which works asymptotically) to write
\begin{equation}
\left[\frac{e^{-\mathcal{R}(\rho)}}{A'(\rho) + \mathcal{R}'(\rho)}\right]^{d-1} \sim \ell^{d-1},\ \ \rho \to \infty.
\end{equation}
Substituting this into the $\rho$-dependent factor in \eqref{sphereafunction1} produces the correct coefficient.

\section{Constructing scalar-AdS flows}\label{app:scalarNums}

In this appendix, we discuss the general numerical procedure by which we construct scalar-induced flows in AdS gravity (setting the curvature scale $\ell = 1$). We solve for classically backreacting black holes of the form
\begin{equation}
ds^2 = \frac{1}{r^2}\left[-e^{-\chi(r)}F(r) dt^2 + \frac{dr^2}{F(r)} + d\Omega_{d-1}^2\right],\ \ \Phi = \phi(r),\label{lorentzAns}
\end{equation}
for which the Einstein + scalar equations of motion \eqref{einsteinScalar}--\eqref{kleinGordon} take the form
\begin{align}
&\phi'' + \left(\dfrac{F'}{F} - \dfrac{d-1}{r} - \dfrac{\chi'}{2}\right)\phi' + \dfrac{\Delta(d-\Delta)}{r^2 F}\phi = 0,\label{eomscd1}\\
&\chi' - \dfrac{2F'}{F} - \dfrac{\Delta(d-\Delta) \phi^2}{(d-1) r F} - \dfrac{2d}{rF} + \dfrac{2d}{r} - \dfrac{2(d-2) r}{F}  = 0,\label{eomscd2}\\
&\chi' -  \dfrac{r}{d-1}(\phi')^2 = 0,\label{eomscd3}
\end{align}
The equations of motion are highly nonlinear, which is why we resort to numerics in the first place. We specifically solve for the metric functions using a shooting method, just as was done for the planar Lorentzian black hole \cite{Frenkel:2020ysx,Caceres:2021fuw}. To construct our black-hole solutions, we shoot from a finite radius taken to be the horizon by setting the blackening function to $0$. More specifically, at some $r = r_{\text{h}}$, we expand each of the metric functions $\{F,\chi,\phi\}$ as
\begin{align}
F(r) &= F_{\text{h}}^{(0)} + F_{\text{h}}^{(1)}(r-r_{\text{h}}) + O[(r-r_{\text{h}})^2],\label{fexp1}\\
\chi(r) &= \chi_{\text{h}}^{(0)} + \chi_{\text{h}}^{(1)}(r-r_{\text{h}}) + O[(r-r_{\text{h}})^2],\\
\phi(r) &= \phi_{\text{h}}^{(0)} + \phi_{\text{h}}^{(1)}(r-r_{\text{h}}) + O[(r-r_{\text{h}})^2].\label{phiexp1}
\end{align}
The equations of motion \eqref{eomscd1}--\eqref{eomscd3} [multiplying the first two by $F(r)$] can then be expanded around $r = r_{\text{h}}$:
\begin{align}
\begin{bmatrix*}[l]
2\phi_{\text{h}}^{(2)}F_{\text{h}}^{(0)}  + \dfrac{\Delta(d-\Delta)}{r_{\text{h}}^2}\phi_{\text{h}}^{(0)}\vspace{0.2cm}\\
\quad+ \left(F_{\text{h}}^{(1)} - \dfrac{d-1}{r_{\text{h}}}F_{\text{h}}^{(0)} - \dfrac{\chi_{\text{h}}^{(1)}F_{\text{h}}^{(0)}}{2}\right)\phi_{\text{h}}^{(1)}\end{bmatrix*}(r-r_{\text{h}})^0 + O(r-r_{\text{h}}) &= 0,\label{const1}\\
\begin{bmatrix*}[l]
- \dfrac{\Delta(d-\Delta)}{(d-1)r_{\text{h}}}\left(\phi_{\text{h}}^{(0)}\right)^2
+ F_{\text{h}}^{(0)}\left(\chi_{\text{h}}^{(1)} + \dfrac{2d}{r_{\text{h}}}\right)\vspace{0.2cm}\\
\quad- 2F_{\text{h}}^{(1)}\left(1 + \dfrac{d}{r_{\text{h}} F_{\text{h}}^{(1)}} + \dfrac{(d-2)r_{\text{h}}}{F_{\text{h}}^{(1)}}\right)\end{bmatrix*} (r-r_{\text{h}})^0 + O(r-r_{\text{h}}) &= 0,\label{const2}\\
\left[\chi_{\text{h}}^{(1)} - \frac{r_{\text{h}}}{d-1}\left(\phi_{\text{h}}^{(1)}\right)^2\right](r-r_{\text{h}})^0 + O(r-r_{\text{h}}) &= 0.\label{const3}
\end{align}
At zeroth order, there are six parameters $\{\phi_{\text{h}}^{(0)},\phi_{\text{h}}^{(1)},\phi_{\text{h}}^{(2)},F_{\text{h}}^{(0)},F_{\text{h}}^{(1)},\chi_{\text{h}}^{(1)}\}$ and three constraints. As we go up order-by-order simultaneously in all three equations of motion, we obtain additional parameters but a matching number of constraints. Distinct solutions are thus specified by three free parameters in the aforementioned list of six.

Note that we have glossed over one of the coefficients $\chi_{\text{h}}^{(0)}$. This is not constrained by the equations of motion, so $\chi$ can only be solved up to an overall constant. However, we have the physical requirement that the metric be asymptotically AdS as $r \to 0$. Thus, the constant term of $\chi$ is that for which $\chi(0) = 0$. In our numerics however, we first set
\begin{equation}
\chi_{\text{h}}^{(0)} = 0,
\end{equation}
and then shift the output function by a constant in order to achieve $\chi(0) = 0$.

Now, we discuss the particular solutions of physical interest. As we are assuming the geometry to be a black hole, we solve for the solutions with
\begin{equation}
F_{\text{h}}^{(0)} = 0.
\end{equation}
This makes $\phi_{\text{h}}^{(2)}$ drop out of the zeroth-order constraints, and more generally $\phi_{\text{h}}^{(2+i)}$ drops out of the $i$th-order constraints. Thus, the black-hole solutions are specified by one free parameter. We take this to be $F_{\text{h}}^{(1)}$. The other constrained lowest-order parameters are
\begin{align}
\chi_{\text{h}}^{(1)} &= -\frac{2 (d-\Delta)\Delta}{r_{\text{h}}^2 F_{\text{h}}^{(1)}} \mathcal{T}_d^1\left(r_{\text{h}};F_{\text{h}}^{(1)}\right),\label{coeffX}\\
\phi_{\text{h}}^{(0)} &= \mp \sqrt{-\frac{2r_{\text{h}} F_{\text{h}}^{(1)}(d-1)}{(d-\Delta)\Delta} \mathcal{T}_d^1\left(r_{\text{h}};F_{\text{h}}^{(1)}\right)},\label{coeffp1}\\
\phi_{\text{h}}^{(1)} &= \pm \sqrt{-\frac{2 (d-1)(d-\Delta)\Delta}{r_{\text{h}}^3 F_{\text{h}}^{(1)}} \mathcal{T}_d^1\left(r_{\text{h}};F_{\text{h}}^{(1)}\right)},\label{coeffP}
\end{align}
where we have defined something which we will call the \textit{topological factor} from \eqref{const2},
\begin{equation}
\mathcal{T}_d^k\left(r_{\text{h}};F_{\text{h}}^{(1)}\right) = 1 + \frac{d}{r_{\text{h}} F_{\text{h}}^{(1)}} + \frac{k(d-2) r_{\text{h}}}{F_{\text{h}}^{(1)}}.\label{funcTop}
\end{equation}
This factor captures the topological dependence of the metric functions. Recall that $k$ labels the spatial topology of the horizon \eqref{topk}. For $k = 1$ (spherical topology), we obtain the equations above. For $k = 0$ (planar topology), we get the equations used in \cite{Frenkel:2020ysx,Caceres:2021fuw}.

With these coefficients in hand, we perform a ``two-sided" shooting. First, we fix a value for $r_{\text{h}}$. As we assume the scalar $\phi$ to be real while $F_{\text{h}}^{(1)} < 0$, we require that the topological factor be nonnegative based on the expressions \eqref{coeffp1} and \eqref{coeffP}. This implies
\begin{equation}
F_{\text{h}}^{(1)} \le -\frac{d}{r_{\text{h}}} - (d-2)r_{\text{h}}.
\end{equation}
For fixed $r_{\text{h}}$, we also fix an array of values for $F_{\text{h}}^{(1)}$ consistent with this bound. Note the value of $F_{\text{h}}^{(1)}$ saturating this bound corresponds to a black hole solving the vacuum equations of motion (with $\phi = 0$), whereas other values of $F_{\text{h}}^{(1)}$ label backreacting solutions.

Given some $r_{\text{h}}$ and $F_{\text{h}}^{(1)}$, we set initial values for the fields $\{F,\chi,\phi\}$ at both $r = r_{\text{h}} - \epsilon$ and $r = r_{\text{h}} + \epsilon$, where $\epsilon$ may be freely chosen so long as $\epsilon \ll r_{\text{h}}$. Since $\epsilon$ is small, we can approximate the fields by truncating the series \eqref{fexp1}--\eqref{phiexp1} at linear order in $\epsilon$. We then compute the asymptotic value of $\chi$ and subtract this number in order to have AdS asymptotics, namely $\chi(0)=0$.

We take this approach over a large range of $r_{\text{h}}$ and $F_{\text{h}}^{(1)}$. This yields a matrix of solutions. We then extract matrices consisting of two different pieces of data: the horizon temperature $T_{\text{h}}$ \eqref{hawkTemp2} and the boundary deformation $\phi_0$ \eqref{phibdryexp}. From here, we have the flexibility to filter out solutions which do not match some sort of selection criterion. For instance, we may only want the solutions for which $\phi_0 \approx 100$ (up to a prescribed uncertainty). This significantly lessens the required computation times for more resource-intensive tasks.

\section{The entropy functional for caps} \label{app:entropyCaps}

Let us briefly set the stage for a possible higher-dimensional ($d > 2$) extension to our exploration of entanglement plateaux in scalar flows. We also point out the mathematical reason why $d = 2$ is simple; the punchline is that the area functional has a ``constant of motion" at $d = 2$.

We pick a parameterization of the $S^{d-1}$ directions in the black hole ansatz such that the metric takes the form
\begin{equation}
ds^2 = \frac{\ell^2}{r^2}\left[-e^{-\chi(r)} F(r) dt^2 + \frac{dr^2}{F(r)} + \ell^2 d\theta_1^2 + \ell^2 \sum_{i=2}^{d-1} \left(\prod_{j=1}^{i-1} \sin^2 \theta_j\right)d\theta_i^2 \right].\label{metbhCap}
\end{equation}
We identify $\theta_1$ as a polar angle ($\theta_1 \sim \theta_1 + 2\pi$). A cap $\mathcal{R}$ on the boundary $r = 0$ is then defined as
\begin{equation}
t = \text{constant},\ \ \theta_1 \in [-\theta_{\mathcal{R}},\theta_{\mathcal{R}}],\label{bcsCapGen}
\end{equation}
for some $0<\theta_{\mathcal{R}}<\frac{\pi}{2}$. To compute the RT surface for such a cap, we may define a codimension-2 surface on some fixed-$t$ Cauchy slice by $\theta_1 = \theta_1(r)$. From the metric \eqref{metbhCap}, the area functional along this surface is
\begin{equation}
\mathcal{A} = \ell^{d-1}  \int \frac{dr}{r^{d-1}} \left[\sin\theta_1(r)\right]^{d-2} \sqrt{\frac{1}{F(r)} + \theta_1'(r)^2}.
\end{equation}
If $d = 2$, then the $[\sin\theta_1(r)]^{d-2}$ factor in the integrand is simply $1$, so the partial derivative of the integrand with respect to $\theta_1$ is simply $0$. Applying the Euler--Lagrange equations thus yields a first-order differential equation, and each extremal surface with boundary conditions \eqref{bcsCapGen} is labeled by a single parameter. However, for $d > 2$, we get a second-order differential equation. While solvable, this also yields a two-parameter family of extremal surfaces, so computing the minimal-area surface is much more difficult.

\end{appendices}
\vfill\pagebreak

\bibliographystyle{jhep}
\bibliography{references.bib}
\end{document}

%% file: figs/roundEntanglement.tex
\begin{figure}
\centering
\begin{tikzpicture}
\draw[-] (0,0) circle (2);
\draw[-] (0,0) circle (0.5);

\draw[-,very thick,red!60] (2,0) arc (0:45:2);
\draw[-,very thick,red!60] (2,0) arc (0:-45:2);

\draw[-,thick,black!30!green] (2/1.414,2/1.414) .. controls (2/1.414,1.5/1.414) and (1.5/1.414,1.5/1.414) .. (1,0.25) ..controls (1-0.0025,0.2) .. (0.99,0) .. controls (1-0.0025,-0.2) .. (1,-0.25) .. controls  (1.5/1.414,-1.5/1.414) and (2/1.414,-1.5/1.414) .. (2/1.414,-2/1.414);

\draw[-,thick,blue!60] (2/1.414,2/1.414) .. controls (2/1.414,1.5/1.414) and (0,1.45) .. (-0.675,0.25) .. controls (-0.73,0.125) and (-0.73,-0.125) .. (-0.675,-0.25) .. controls (0,-1.45) and (2/1.414,-1.5/1.414) .. (2/1.414,-2/1.414);

\draw[-,thick,blue!60,fill=black,pattern=north west lines] (0,0) circle (0.5);

\node[red!60] at (2/1.414,2/1.414) {$\bullet$};
\node[red!60] at (2/1.414,-2/1.414) {$\bullet$};
\node at (2/1.414,2/1.414) {$\circ$};
\node at (2/1.414,-2/1.414) {$\circ$};

\draw[->,very thick,red!60] (2.3,0) arc (0:45:2.3);
\draw[->,very thick,red!60] (2.3,0) arc (0:-45:2.3);
\node[red!60] at (2.5,0) {$\mathcal{R}$};
\node[white] at (-2.5,0) {$\mathcal{R}$};

\node[black!30!green] at (1.3,0) {$\Gamma_{\text{C}}$};
\node[blue!60] at (-1.125,0) {$\Gamma_{\text{D}}$};
\end{tikzpicture}
\caption{A fixed-time slice of the round black hole, with the RT candidates corresponding to a boundary region $\mathcal{R}$. Generically, the true minimal RT surface may either be connected ($\Gamma_{\text{C}}$) or disconnected ($\Gamma_{\text{D}}$). Note that $\Gamma_{\text{D}}$ itself has two connected components---a piece $\overline{\Gamma}_{\text{D}}$ that wraps around the black hole horizon and the horizon itself.}
\label{figs:roundEntanglement}
\end{figure}

%% file: figs/round2pt.tex
\begin{figure}
\centering
\begin{tikzpicture}
\draw[-] (0,0) circle (2);
\draw[-] (0,0) circle (0.5);


\draw[-,thick,yellow!60!black] (2/1.414,2/1.414) to  (0.5/1.414,0.5/1.414);
\draw[-,thick,yellow!60!black] (2/1.414,-2/1.414) to  (0.5/1.414,-0.5/1.414);

\draw[-,thick,black!30!green] (2/1.414,2/1.414) .. controls (2/1.414,1.5/1.414) and (1.5/1.414,1.5/1.414) .. (1,0.25) ..controls (1-0.0025,0.2) .. (0.99,0) .. controls (1-0.0025,-0.2) .. (1,-0.25) .. controls  (1.5/1.414,-1.5/1.414) and (2/1.414,-1.5/1.414) .. (2/1.414,-2/1.414);

\draw[-,thick,blue!60] (2/1.414,2/1.414) .. controls (2/1.414,1.5/1.414) and (0,1.45) .. (-0.675,0.25) .. controls (-0.73,0.125) and (-0.73,-0.125) .. (-0.675,-0.25) .. controls (0,-1.45) and (2/1.414,-1.5/1.414) .. (2/1.414,-2/1.414);

\node[yellow!60!black] at (1/1.414-0.2,1/1.414+0.075) {D};

\draw[-,thick,fill=black,pattern=north west lines] (0,0) circle (0.5);

\node[red!60] at (2/1.414,2/1.414) {$\bullet$};
\node[red!60] at (2/1.414,-2/1.414) {$\bullet$};
\node at (2/1.414,2/1.414) {$\circ$};
\node at (2/1.414,-2/1.414) {$\circ$};

\node[red] at (2/1.414+0.35,2/1.414+0.15) {$\mathcal{O}_{\text{H}}$};
\node[red] at (2/1.414+0.35,-2/1.414-0.15) {$\mathcal{O}_{\text{H}}$};

\node[white] at (-2/1.414-0.35,2/1.414+0.15) {$\mathcal{O}_{\text{H}}$};
\node[white] at (-2/1.414-0.35,-2/1.414-0.15) {$\mathcal{O}_{\text{H}}$};


\node[black!30!green] at (1.3,0) {C1};
\node[blue!60] at (-1.125,0) {C2};
\end{tikzpicture}
\caption{A fixed-time slice of the round black hole (in Lorentzian signature), with the two connected geodesics (C1 and C2) and the disconnected geodesic (D) which contribute to the heavy thermal two-point function $\expval{\mathcal{O}_{\text{H}}\mathcal{O}_{\text{H}}}$. The shortest geodesic in this black hole background dominates the Euclidean scalar two-point function computed in a thermal state. As we are actually computing this two-point function in Euclidean signature, the disconnected geodesic ends at the horizon.}
\label{figs:round2pt}
\end{figure}